\documentclass{article}

\usepackage[T1]{fontenc} 
\usepackage[normalem]{ulem} 

\usepackage[usenames,dvipsnames,svgnames,x11names]{xcolor}
\newcommand{\Note}[1]{\textcolor{red}{#1}} 
\newcommand{\ysnoted}[1]{ {\textcolor{green} { ***TODO Later: #1 }}} 

\newcommand{\modc}[1]{{\textcolor{blue}{#1}}}
\newcommand{\addc}[1]{{\textcolor{teal}{#1}}}
\newcommand{\delc}[1]{ {\textcolor{gray} {\sout{#1}} }}
\newcommand{\modcr}[1]{{\textcolor{blue}{#1}}}
\newcommand{\addcr}[1]{{\textcolor{teal}{#1}}}
\newcommand{\delcr}[1]{ {\textcolor{gray} {\sout{#1}} }}
\newcommand{\modcrr}[1]{{\textcolor{blue}{#1}}}
\newcommand{\addcrr}[1]{{\textcolor{teal}{#1}}}

\renewcommand{\modc}[1]{#1} 
\renewcommand{\addc}[1]{#1} 
\renewcommand{\delc}[1]{} 
\renewcommand{\modcr}[1]{#1} 
\renewcommand{\addcr}[1]{#1} 
\renewcommand{\delcr}[1]{} 
\renewcommand{\modcrr}[1]{#1} 
\renewcommand{\addcrr}[1]{#1} 

\renewcommand{\ysnoted}[1]{} 

\usepackage{listings}
\lstdefinelanguage{XML}
{
basicstyle=\ttfamily\footnotesize,
  morestring=[b]",
  moredelim=[s][\bfseries\color{Maroon}]{<}{\ },
  moredelim=[s][\bfseries\color{Maroon}]{</}{>},
  moredelim=[l][\bfseries\color{Maroon}]{/>},
  moredelim=[l][\bfseries\color{Maroon}]{>},
  morecomment=[s]{<?}{?>},
  morecomment=[s]{<!--}{-->},
  commentstyle=\color{gray},
  stringstyle=\color{blue},
  identifierstyle=\color{red}
}
\usepackage{moreverb}

\usepackage[nounderscore]{syntax}

\usepackage[cmex10]{amsmath}
\usepackage{amssymb}
\usepackage{mathtools}
\usepackage{amsthm}
\usepackage{amsfonts}

\usepackage{gensymb}

\usepackage{subfig} 
\usepackage[export]{adjustbox}

\usepackage{algorithmicx}
\usepackage{algpseudocode}
\usepackage[ruled]{algorithm}
\definecolor{light-gray}{gray}{0.75}
\definecolor{dark-gray}{gray}{0.5}
\algrenewcommand{\algorithmiccomment}[1]{\hskip3em{{\footnotesize \textcolor{light-gray}{$\blacktriangleright$}~\textcolor{dark-gray}{#1}}}}
\usepackage{multirow} 
\usepackage{rotating} 
\usepackage{booktabs} 
\usepackage{colortbl} 
\usepackage{tablefootnote} 

\usepackage[pdftex,colorlinks=true,urlcolor=blue,citecolor=blue]{hyperref}

\usepackage{xspace}

\usepackage{enumitem}

\usepackage{balance}

\newcommand{\DS}{DEMS\xspace}
\newcommand{\GS}{GEMS\xspace}

\newcommand{\HV}{\textsf{HV}\xspace}
\newcommand{\MD}{\textsf{MD}\xspace}

\newcommand{\BP}{\textsf{BP}\xspace}

\newcommand{\CD}{\textsf{CD}\xspace}

\newcommand{\DEO}{\textsf{DEO}\xspace}
\newcommand{\DEV}{\textsf{DEV}\xspace}

\newcommand{\SJFe}{SJF\xspace}
\newcommand{\HPFe}{HPF\xspace}
\newcommand{\HUFe}{HUF\xspace}
\newcommand{\CLD}{CLD\xspace}
\newcommand{\EC}{E+C\xspace}
\newcommand{\EDFe}{EDF\xspace}
\newcommand{\EDFcm}{DEM\xspace}

\newcommand{\twoBP}{2D-P\xspace}
\newcommand{\twoBA}{2D-A\xspace}
\newcommand{\threeBP}{3D-P\xspace}
\newcommand{\threeBA}{3D-A\xspace}
\newcommand{\fourBP}{4D-P\xspace}
\newcommand{\fourBA}{4D-A\xspace}

\usepackage{blindtext}

\usepackage{tikz}

\begin{document}
\date{}


\title{
Adaptive Heuristics for Scheduling DNN Inferencing on Edge and Cloud for Personalized UAV Fleets
}

\author{Suman Raj, Radhika Mittal, Harshil Gupta and Yogesh Simmhan\\
Department of Computational and Data Sciences, \\Indian Institute of Science, Bangalore 560012 India\\
Email: \{sumanraj, simmhan\}@iisc.ac.in
}

\maketitle
\thispagestyle{plain}
\pagestyle{plain}

\begin{abstract}

Drone fleets with onboard cameras coupled with computer vision and DNN inferencing models can support diverse applications, from package deliveries to disaster monitoring. One such novel domain is for one or more ``buddy'' drones to assist Visually Impaired People (VIPs) lead an active lifestyle. Video inferencing tasks from such drones can help both navigate the drone and provide situation awareness to the VIP, and hence have strict execution deadlines. These tasks can execute either on an accelerated edge like Nvidia Jetson linked to the drone, or on a cloud INFerencing-as-a-Service (INFaaS). However, making this decision is a challenge given the latency and cost trade-offs across a stream of deadline-sensitive tasks, in the presence of network and/or compute variability.
We propose a deadline-driven heuristic, \DS-A, to schedule diverse DNN tasks generated continuously to perform inferencing over video segments generated by multiple drones linked to an edge, with the option to execute on the cloud. We use strategies like task dropping, work stealing and migration, and dynamic adaptation to cloud variability, to fully utilize the captive edge with intelligent offloading to the cloud, \addc{and guarantee a Quality of Service (QoS), \modcr{\textit{i.e.}}} maximize the utility and the number of tasks completed. \addc{We also introduce an additional Quality of Experience (QoE) metric 
useful to the assistive drone domain, which values the frequency of success for task types to ensure the responsiveness and reliability of the VIP application. We extend our DEMS solution to GEMS to solve this.} 
\modc{We evaluate these strategies, using (i) an emulated setup of a fleet of over $80$ drones supporting over $25$ VIPs, with real DNN models executing on pre-recorded drone video streams, using Jetson Nano edges and AWS Lambda cloud functions,} \addc{and (ii) a real-world setup of a Tello drone and a Jetson Orin Nano edge accelerator executing a subset of the DNN models on live video feeds and generating drone commands to follow a VIP in real-time.}
\modc{The detailed comparative emulation study shows that our strategies have a task completion rate of up to $88\%$, up to $2.7\times$ higher QoS utility compared to the baselines, a further $16\%$ higher QoS utility while adapting to network variability,} \addc{and up to $75\%$ higher QoE utility. Our practical validation using real drones exhibits task completion of up to $87\%$ for \GS and $33\%$ higher total utility of \GS compared to edge-only and achieves the smoothest trajectory with minimum jerk and lowest yaw error.}
\end{abstract}


\vspace{-0.3cm}
\section{Introduction} 
\vspace{-0.2cm}

\modcr{According to the World Health Organization (WHO), globally, over 2.2 billion people suffer from moderate or severe visual impairment worldwide, of whom 36 million are blind~\cite{whoStats}. This has a severe impact on their quality of life and can lead to social isolation, difficulty in mobility, and a higher risk of falls. Towards this, we are developing a novel and socially beneficial application to leverage ``buddy'' drones to help \textit{Visually Impaired People (VIPs)} lead an independent, active, and safe outdoor lifestyle~\cite{suman2023chi}.} Unmanned Aerial Vehicles (UAVs), or drones, are increasingly becoming a flexible mobility and observation platform. Progressive regulations by the US FAA, Europe's EASA, and India's DGCA are bringing industry and researchers to examine disruptive use cases for drones~\cite{faa,easa,dgca}. Fleets of drones are starting to be used in cities and towns for urban safety, transport and infrastructure monitoring~\cite{9128519}, logistics and delivery~\cite{benarbia2021literature}, and even to assist \addc{in combating COVID-19 pandemic~\cite{KUMAR20211}, in healthcare applications~\cite{hiebert2020application}} or navigating the visually challenged~\cite{avila2015dronenavigator,avila2017dronenavigator}.

\delcr{Drones come in a variety of form factors. Nano quad-copters, which weigh $<100g$, have onboard cameras, and use data links over WiFi to a base station, are popular and cheap. \textit{e.g.}, the DJI Tello drone~\footnote{\href{https://www.ryzerobotics.com/tello}{Ryze Tello drones specifications}} we use costs under US\$100. Drones can also access GPU-accelerated edge devices, such as Nvidia Jetson, either on-board or through a wireless link to the base station.}

\delcr{\modcr{\textit{e.g.}}, a Jetson Orin Nano has $1024$ Ampere CUDA cores, a $6$-core Arm Cortex-A78AE CPU, $8$~GB RAM shared by CPU and GPU, consumes $7$--$15$~W of power, measures just $100 \times 79$~mm~\cite{orin-tech-specs} -- slightly larger than a smartphone, and costs about US\$499.}
\delcr{Several drones can cooperatively complete a task, connecting to the same base station and edge accelerator.}

%
\subsection{\modcr{Motivation as a Drones-Assistive Technology}}
\modcr{Drones come in a variety of form factors. Nano quad-copters, which weigh $<100g$, have onboard cameras, and use data links over WiFi to a base station, are popular and cheap. \textit{e.g.}, the DJI Tello drone~\footnote{\href{https://www.ryzerobotics.com/tello}{Ryze Tello drones specifications}} we use costs under US\$100. Drones can also access GPU-accelerated edge devices, such as Nvidia Jetson, either on-board or through a wireless link to the base station. \textit{E.g.}, a Jetson Orin Nano has $1024$ Ampere CUDA cores, a $6$-core Arm Cortex-A78AE CPU, $8$~GB RAM shared by CPU and GPU, consumes $7$--$15$~W of power, measures just $100 \times 79$~mm~\cite{orin-tech-specs} -- slightly larger than a smartphone, and costs about US\$499. While numerous applications of drones to commercial delivery and security exist, we are exploring the use of drones as an assistive technology~\cite{kuriakose2022tools,al2016exploring}. One or more autonomous drones which coordinate among themselves can provide a comprehensive view of the environment and adjust to the mobility of the VIP while navigating city streets. The drones can also be dispatched to get, say, a first aid kit in an emergency~\cite{drone-first-aid}.}

\delcr{According to the World Health Organization (WHO), globally, over 2.2 billion people suffer from moderate or severe visual impairment worldwide, of whom 36 million are blind~\cite{whoStats}. This has a severe impact on their quality of life and can lead to social isolation, difficulty in mobility, and a higher risk of falls. Towards this, we are developing a novel and socially beneficial application to leverage ``buddy'' drones to help \textit{Visually Impaired People (VIPs)} lead an independent, active, and safe outdoor lifestyle~\cite{suman2023chi}.} 

\delcr{Besides traditional white canes and guide dogs, technology interventions for supporting VIPs include sensor-based walking assists~\cite{islam2019developing}, smart canes~\cite{weWalk}, and AI-powered backpacks~\cite{aiPoweredBackpack} that analyze the environment using sensors to warns users. However, the VIP has to physically carry them, curtailing their autonomy and also limiting the field of view of the camera for sensing. One or more autonomous drones which coordinate among themselves can provide a comprehensive view of the environment and adjust to the mobility of the VIP while navigating city streets. The drones can also be dispatched to get, say, a first aid kit in an emergency~\cite{drone-first-aid}.}

Much of this is possible thanks to advances in computer vision or Deep Neural Network (DNN) models that can automatically analyze video and other sensor data collected from drones. The output of such model inferencing can support real-time applications for the VIP.
\modcr{\textit{e.g.}}, multiple buddy drones supporting a VIP can perform DNN inferencing over their video streams to (1) recognize, track and autonomously follow the VIP in real-time; (2) identify and warn them about hazards rapidly, such as an obstruction or a crowd not wearing masks during the pandemic; and (3) notify them if they are near a point of interest, such as a park or a pharmacy.
In future, this can be generalized to an app platform ecosystem to design and build many such innovative drone-based applications, with collaboration across fleets of buddy drones and edge devices of different VIPs.

\subsection{Challenges and Gaps} 

\begin{figure}[t!]
\vspace{-0.2in}
\centering
  \subfloat[\modc{Container mimicking Jetson Nano Edge}]{
  \quad  \includegraphics[width=0.45\columnwidth]{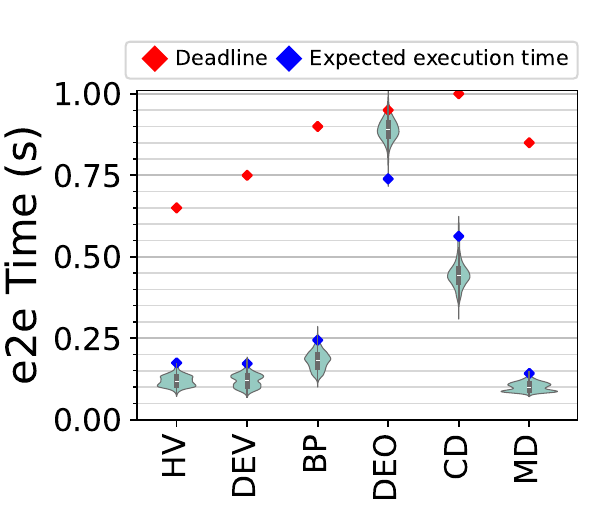} \quad
   \label{fig:edge_benchmark_1drone}
  }
  \subfloat[\modc{AWS Lambda FaaS}]{
   \includegraphics[width=0.45\columnwidth]{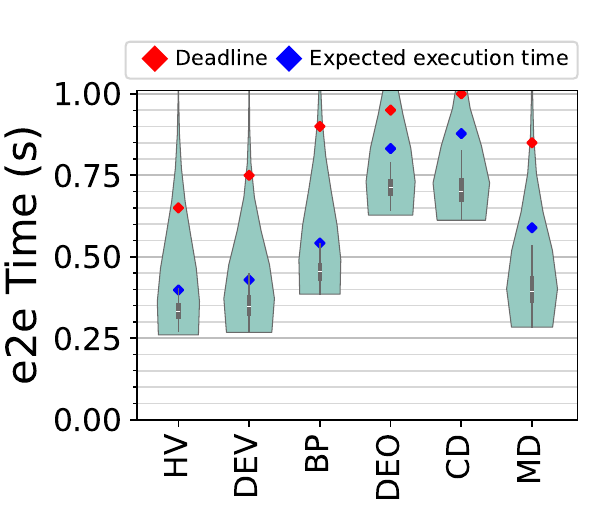}
    \label{fig:lambda_benchmark_1drone}
  }
\vspace{-0.07in}
\caption{\modc{Model Inferencing Time Distribution for 6 DNNs.} }
\label{fig:bm:compute}
\vspace{-0.2in}
\end{figure}

Supporting such a \textit{cyber-physical-social application} is complex and interdisciplinary~\cite{ZENG20201028}. Besides challenges in computer vision models and control systems for autonomous navigation, this also requires robust means for real-time inferencing. \textit{Multiple DNN models} need to execute over a \textit{stream of video segments} that is continuously collected by the drones of the VIP and their surroundings, and drive decisions with diverse priorities.
The tasks may have hard \textit{deadlines} based on their time-sensitivity and different \textit{benefits} based on their importance. \modc{For instance, using vision algorithms to track the VIP to help the drone navigate and follow them may have a short deadline and medium benefit; estimating the distances to nearby vehicles and alerting the VIP about it may have a medium deadline and high benefit; while visually detecting a point of interest like a park bench may have a relaxed deadline with low benefit.} \modcrr{Assigning deadlines and benefits for a task is application-specific. For instance, estimating distances to nearby vehicles for a VIP has a moderate deadline when the distances are pre-estimated over previous frames but may require a shorter deadline under fast-moving scenarios, such as oncoming traffic, to ensure timely safety interventions.} 
\addc{These should be customizable to a particular domain and a VIP.}

%
DNN inferencing is compute-heavy and requires accelerators for timely processing. The base station which receives the video streams from the drones can perform these DNN tasks on its co-located \textit{edge accelerator}, or send it to an \textit{INFe\-ren\-cing-as-a-Service (INFaaS)} on the cloud, based on intelligent trade-offs. INFaaS comes in various flavors, ranging from pre-defined APIs for voice-to-text from cloud providers ({\modcr{\textit{e.g.}}, Amazon Transcribe) to custom DNN models deployed by developers using Functions-as-a-Service (FaaS) or as services hosted on (accelerated) VMs. We adopt the FaaS design to wrap our inferencing models due to its rapid elasticity and reasonably reliable execution times.

\delc{Low-end} Edge GPU accelerators like Nvidia Jetsons 
\delc{have certain differences from desktop/server GPUs, {\modcr{\textit{e.g.}}}, not supporting task parallelism and having shared memory across CPU and GPU~\cite{10.1145/3570604}. The execution \Note{duration} for the same task can be different on the edge or the cloud, as is the parallelism that can be exploited.}
\addc{complement the cloud resources through local inferencing on captive devices with lower amortized execution cost compared to FaaS~\cite{10.1145/3570604}.}
\delc{The difference between a task's benefit, if executed within its deadline, and its execution cost, is its \textit{utility}. This can be negative for a task that is executed on the cloud.}
While the DNN inferencing duration on the \textit{cloud} is faster, the end-to-end execution time may be higher due to the network transfer time to send the video segments from the edge. This is exacerbated due to the mobility of drones and the use of cellular networks, causing high variability in network latency and bandwidth.
\delc{and hence end-to-end cloud inferencing times. Responding to this variability when scheduling cloud tasks is critical.}%
The \textit{edge} has a slower accelerator and finite capacity but no network variability. 
\delc{While the captive edge has low to no cost, FaaS has a pay-as-you-go cost per function invocation that depends on the inferencing time and the memory usage.}

\begin{figure}[t!]
\vspace{-0.1in}
\centering
  \subfloat[WAN--Cloud Ping]{
    \includegraphics[width=0.255\columnwidth]{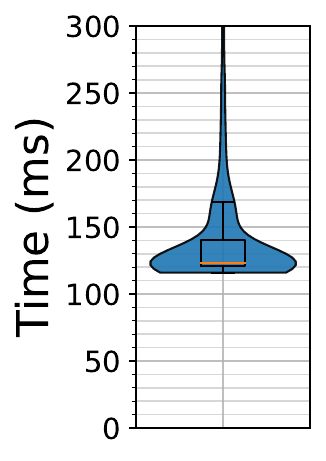}
   \label{fig:bm:nw:lan-lat}
  }%
  \subfloat[WAN--Cloud BW]{
    \includegraphics[width=0.255\columnwidth]{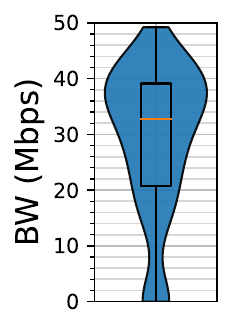}
    \label{fig:bm:nw:lan-bw}
  }%
  \subfloat[Mobile device--Cloud BW (NS3)]{
    \includegraphics[width=0.46\columnwidth]{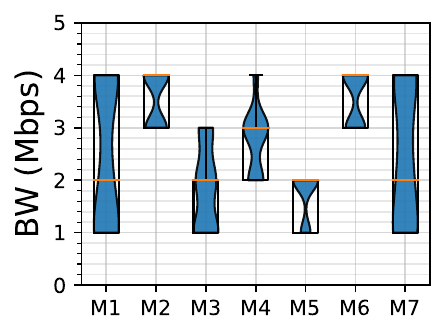}
    \label{fig:bm:nw:wan-bw}
  }
\vspace{-0.07in}
\caption{\modc{Network Characteristics for WAN (real) and Cellular (simulated).}}
\label{fig:bm:nw}
\vspace{-0.2in}
\end{figure}

\modc{We empirically characterize this by benchmarking the inferencing time and network variability of edge and cloud inferencing for six different DNNs. We run the models on an \textit{(edge) container} deployed on a local server that mimics the performance of a Jetson Nano, and on \textit{AWS Lambda FaaS}, hosted at the nearest \texttt{ap-south-1} data center (Asia Pacific, Mumbai). These DNNs are more fully described in Sec.~\ref{sec:results}. The clients are on our campus network, serving as a proxy for the drones sending inference requests, and co-located with the edge container. \addcrr{Since the edge scheduler and executor are co-located on the same device, the setup inherently avoids external network delays, resulting in negligible network variability. This is further described in Sec.~\ref{sec:system-model}.}
We make $\approx2k$ calls to each model on the edge and the cloud.}

\modc{Fig.~\ref{fig:bm:compute} reports the end-to-end inferencing time. 
The edge inferencing time is lower and tighter than the AWS FaaS time, explained by the WAN network variability when connecting to the cloud. This is confirmed when we separately measure the latency and bandwidth from the client to cloud in Figs.~\ref{fig:bm:nw:lan-lat} and~\ref{fig:bm:nw:lan-bw}), which show a long tail (ping) and high divergence (bandwidth).
\modcr{This variability increases when we simulate the mobility of 7 drones on our campus using the SUMO simulator, while leveraging a 4G cellular network in the NS3 simulator.}
Fig.~\ref{fig:bm:nw:wan-bw} shows even more diversity across devices due to the mobility.}

These benchmarks motivate the need to adapt to such network dynamism when scheduling for our drone application. 
\delc{When coupled with the need to meet strict deadlines, this poses interesting challenges for the base station to schedule a stream of tasks on the edge and cloud to maximize the utility gained \textit{per VIP} and also increase the on-time task completion.}
\addc{Such scheduling decisions should also go beyond just achieving \textit{Quality of Service (QoS)} guarantees of meeting the deadlines for individual tasks, to assuring the \textit{Quality of Experience (QoE)} exposed to the VIP application. \modcr{\textit{e.g.}}, it may be acceptable to miss the deadline for a few video frames by a DNN tracking the VIP as long as, say, at least 10 out 15 frames in a 15~second window are inferred on time to help the drone navigate smoothly. But a DNN detecting hazards from moving vehicles may need timely detections within a narrower window, \modcr{\textit{e.g.}}, 2 out of 3 frames in a 3~second window, to warn users. This motivates the need for higher-order QoE metrics that are usable by the VIP application.}

Existing scheduling algorithms do not adequately consider the use of both edge and cloud~\cite{7300228}, diversity in DNN inferencing tasks~\cite{8685783}, tight deadlines for tasks~\cite{guo2019uav}, network and execution time variability on the cloud~\cite{guo2019uav}, FaaS and edge pricing models~\cite{khochare:ton:2024}, \addc{guarantees task completion rate~\cite{10171496}} and/or multiple continuous task streams jointly scheduled by one edge~\cite{8685783}.
\addc{QoE has been examined as a domain-facing metric to provide service accessibility~\cite{MAHMUD2019190} and real-time inferencing~\cite{9723632}, but not for autonomous vehicles to meet domain-specified metrics, \modcr{\textit{e.g.}}, to achieve a smooth trajectory for the drones.}  

\modc{In this article, our proposed \DS algorithm finely balances both the objectives of on-time completion of tasks and utility maximization, while \GS helps maintain a task completion rate set within a window, as determined by the user and the application.} Our heuristics, while leveraging existing concepts, are specifically tuned toward the VIP application and, more generally, to support inferencing for a fleet of drones.

%
\subsection{Contributions}
We make the following specific contributions in this article:
\begin{enumerate}[leftmargin=*,noitemsep]
\item \modc{We propose an \textit{inferencing architecture} to enable a novel \textit{application platform} that uses drones as buddies to support visually impaired people (VIPs), with multiple ``apps'' using video analytics to make decisions and alerts for the VIP and the drone} (\S~\ref{sec:vip},~\ref{sec:app-based-framework}).

\item We use this to motivate and define the problem of \textit{real-time deadline-driven scheduling of DNN inferencing tasks} on the accelerated edge and cloud FaaS to \addc{meet the QoS}, which is to maximize the \addc{QoS} utility and the \addc{on-time} task completion count (\S~\ref{sec:problem}). \addc{We further include a user-facing QoE metric on the task completion rate within a window, which additionally accrues a QoE utility.}

\item We propose a deadline-driven scheduling heuristic \modc{\textit{\DS-A}} for \textit{\uline{d}rones} that incorporates strategies for preemptive dropping of tasks based on \textit{\uline{e}arliest deadline}, their \textit{\uline{m}igration} from edge to cloud, \textit{work \uline{s}tealing} from cloud back to edge, with \textit{\uline{a}daptation} to network variability (\S~\ref{sec:scheduling}).  \addc{This is complemented by an adaptive \textit{\GS} heuristic that \textit{\uline{g}uarantees} the QoE completion rate target set by the apps (\S~\ref{sec:gatekeeper}).} 

\item We \textit{evaluate and compare} our heuristics against baseline scheduling algorithms \delc{that schedule only on edge and on both edge and cloud} for realistic VIP workloads over real drone video feeds. \DS outperforms by completing up to \modc{$88\%$} of tasks, gaining up to \modc{$2.7\times$} higher \modc{QoS} utility. \modc{The DEMS-A variant further improves the QoS utility by \modc{$16\%$} when adapting to network variability.} \addc{GEMS is comparable to DEMS on total utility but has up to $75\%$ higher QoE utility} (\S~\ref{sec:results}).

\item We also validate the practical efficacy of our scheduler for the VIP application using a real drone coupled with Jetson \modc{Orin} Nano for navigation assistance within our campus \addc{(\S~\ref{subsec:hardware-validation})}. \addc{We report domain metrics on the smoothness of the drone's trajectory when navigated using inferencing done by \GS, \DS and the baselines. \GS at $30$~FPS achieves the smoothest trajectory with \textit{minimum jerk} and yaw error.}

\end{enumerate}
\noindent \modc{In addition}, we contrast our work with related literature (\S~\ref{sec:related})\addcr{, discuss legal issues and limitations (\S~\ref{sec:discussion})} and offer our conclusions and future work (\S~\ref{sec:conclusion}).

\addc{An earlier short version~\cite{10171496} of this article motivated the application, defined the first part of the deadline-driven QoS optimization problem, proposed the \DS heuristic to maximise QoS, and reported emulation experiments that compared it against baseline scheduling algorithms.
The current article improves upon it by including QoE as an additional optimization goal and proposes the \GS heuristic to address this (\S~\ref{sec:gatekeeper}), (\S~\ref{subsec:qoe-results}). We also report results using current state-of-the-art DNN models in our workloads.
Further, \modcr{we} integrate these within an end-to-end autonomous VIP navigation solution (\S~\ref{sec:app-based-framework}) and perform \delc{a} real-world validation with a Tello drone, Orin Nano and AWS Lambda functions for real-time experiments to navigate a proxy VIP in our campus, and report domain results (\S~\ref{subsec:hardware-validation}). }

\vspace{-0.3cm}
\section{Related Work}\label{sec:related}
\vspace{-0.2cm}

\subsection{\modcr{Outdoor Assistive Technologies for VIPs}}

Despite active research on navigation support for VIPs, technologies to assist the visually impaired for outdoor navigation are still evolving. \addcr{Besides traditional white canes and guide dogs, technology interventions for supporting VIPs include sensor-based walking assists~\cite{islam2019developing}, smart canes~\cite{weWalk}, and AI-powered backpacks~\cite{aiPoweredBackpack} that analyze the environment using sensors to warns users. However, the VIP has to physically carry them, curtailing their autonomy and also limiting the field of view of the camera for sensing.} Smart guiding glasses~\cite{bai2017smart} use depth cameras to warn users about threats, while Google's Lookout app~\cite{lookout} uses DNN inferencing for simple tasks like identifying currency. Both require users to look/point in a specific direction rather than offer ambient sensing that our drones give.

Others~\cite{avila2015dronenavigator,avila2017dronenavigator,al2016exploring} use lightweight drones to guide VIPs using their distinct sound. But this is limited to pre-defined paths or, in DroneNavigator~\cite{avila2017dronenavigator}, using a bracelet to digitally guide the VIP. They do not use vision-based inferencing like us, which is more flexible, and also motivates the need for real-time analytics on edge and cloud. Nasralla, et al.~\cite{nasralla2019computer} have studied the prospects of using deep learning and computer vision-enabled UAVs for assisting VIPs in a smart city. We uniquely use multiple buddy drones coupled with deadline-aware inferencing for a 360$\degree$ perspective, to help operate in dynamic envi\-ron\-ments. This overcomes the sensing, mobility, and responsiveness gaps.

\subsection{Edge, Cloud and Drone Scheduling}

There has been a significant amount of work on edge, fog and cloud scheduling~\cite{8622052}. \addc{Varshney, et al.~\cite{varshney2020characterizing} provides a detailed survey on the resource capabilities and limitations of the edge, fog and cloud resources, and a taxonomy on scheduling techniques for different applications.}  EDF-PStream~\cite{7300228} presents efficient EDF scheduling for automotive applications. \delcr{OnDisc~\cite{8917749} minimizes the total weighted response time of latency sensitive jobs in heterogeneous edge-clouds.} Dai, et al.~\cite{dai2019scheduling} proposes a task scheduling algorithm for time-constraint tasks in autonomous driving using mobile-edge computing servers. But our application poses unique challenges to deadline-aware and cost-effective scheduling on edge and cloud, in the presence of mobility and network variability, and the inferencing asymmetry between accelerated edge and cloud resources. Our heuristics are tuned to this specialized domain to support the VIP app platform.

\delcr{Yahya, et al.~\cite{yahya2020multi} use multi-criteria decision-making for task coordination on cloud and fog using fuzzy logic. They target streaming data analysis in real-time with varying workload intensities while ensuring high availability and data quality. Others~\cite{li2021cooperative} jointly optimize the cooperative placement and scheduling of services on resource-limited edge-clouds for delay-sensitive requests, while pre-emptively dropping tasks. Neither of these papers considers cloud dynamism or edge mobility, or leverage edge/cloud accelerators. }

\modc{Scheduling on hybrid clouds and 5G networks has also been considered~\cite{ABURUKBA2020539}}.  Dedas~\cite{8941266} proposes an online scheduling algorithm that adopts the idea of EDF along with other algorithms for latency-sensitive applications considering network bandwidth and resource constraints.  Eco~\cite{rao2021eco} proposes a min-cut algorithm to dynamically map micro-services to the edge or cloud to meet application response times in a 5G network. As their edge is part of the 5G base station, their edge cost is higher than the cloud. We use free/cheap edges, and this impacts the scheduling design. Zhang, et al.~\cite{zhang2022deadline} propose DSOTS for a time-sensitive IoT application using edge-cloud collaborative computing. However, they do not take network variability into account. \delcr{Junlong, et al.~\cite{zhou2019cost} optimize the execution cost and makespan for workflows scheduled in hybrid clouds, \addc{while Ghosh, et. al~\cite{10.1145/3140256} propose GA to optimize the energy aware placement of CEP queries and aim to minimize the makespan of the workflow}. But they assume fixed execution times for all tasks and do not explore heterogeneous workloads using a variety of DNN models.}

Task scheduling for autonomous systems has also been explored. Chen, et al.~\cite{8675170} propose a hybrid computing model, UAV-Edge-Cloud to provide powerful resources to support resource-intensive applications and real-time tasks at edge networks.
Postoaca, et al.~\cite{postoaca2020deadline} use a cloud-edge architecture for deadline-constrained face-recognition by robots, similar to our DNN inferencing for drone videos. But they do not have accelerated edges and hence push all inferencing to the cloud. Khochare, et al.~\cite{khochare:ton:2024} propose a co-scheduling problem to both route drones and schedule tasks on-board their edge accelerators within a deadline, but do not use the cloud. We intelligently use both edge and cloud, and maximize task completion and utility.


\subsection{QoS \addc{and QoE}-aware Scheduling}
The placement of DNN inferencing tasks has also gained recent attention. The Kalmia~\cite{fu2022kalmia} QoS-aware framework for scheduling DNN tasks on edge servers splits them into urgent and non-urgent tasks. We allow more flexibility to users by specifying a deadline and a benefit per task. Chen, et al.~\cite{chen2021energy} execute different layers of a DNN inferencing model across edge and cloud, coupled with an energy model, for smart IoT systems with deadline constraints. They assume that the cloud-edge environment has constant performance while our drone mobility and public clouds introduce resource dynamism, which we adapt to. 
\delcr{Various approaches have been studied for SLA agreements for cloud computing for task scheduling~\cite{panda2017sla,NASR2018424} and resource management~\cite{GARCIAGARCIA20141}.}
Recently, SLA-guarantees are also gaining traction in edge-cloud systems~\cite{9996361}, and UAV with edge-cloud~\cite{8675170}. They evaluate their algorithms using simulation and with randomized workload parameter values for their UAV-Edge-Cloud computing model. We instead follow a realistic model based on hardware benchmarking and actual costs.}   

\addc{Quality of Experience (QoE) has emerged as a user-facing metric that complements QoS, which tends to be system-focused. Generally, a user’s QoE increases with the QoS but the former better captures the benefit to the application, especially for user-facing and cyber-physical-social domains~\cite{LAI2020684}. The use of QoE has been examined for scheduling in edge and fog computing.
Li, et al.~\cite{LI201993} aim to improve the user experience through joint placement of data blocks and scheduling of tasks to reduce the response time of tasks completion in an edge computing scenario. \delcr{Similarly, Mahmud et al.~\cite{MAHMUD2019190} propose a QoE-aware application placement policy in fog computing that prioritizes application placement requests based on user expectations while investigating resource availability, proximity, and processing capabilities of fog computational instances.} We not only aim to reduce the task response within a deadline but also ensure a user-defined minimum task completion rate within a time window. Also, our use of cloud to complement the edge helps maintain QoE even when the workload increases, unlike their results which show the QoE decrease for larger workloads.}

\addc{The QoE metric is also used for real-time inferencing and in autonomous systems~\cite{9005248}.
Mao, et al.~\cite{9723632} propose a differentiated QoE scheduler (DQoES) and monitor QoE targets for DNN inferencing using a cloud-only approach. The cloud containers achieve the QoE by vertically \textit{scaling up} the container sizes for varying workload. We instead \textit{scale-out} from edge to the cloud to deal with variable workloads while also minimizing the cloud FaaS costs relative to utility gained. Adapting based on real-time information is a key aspect of QoE critical in scenarios like video streaming by UAVs for firefighting teams~\cite{9714786} and for our VIP use case. Our GEMS heuristic considers the deadline of essential tasks and ensures that the QoE metric of task completion within a window is fulfilled.}

\subsection{\addcr{Summary of Gaps}}
\addcr{Existing VIP navigation aids have limited sensing, while prior drone-based approaches rely on predefined paths or audio guidance without real-time vision inferencing. Our method uniquely integrates multiple buddy drones with deadline-aware inferencing for $360^{\circ}$ perspective. Task scheduling on the edge-cloud continuum often overlooks network variability and cost-effective scheduling. While some UAV-edge-cloud approaches optimize resource use, they either rely entirely on cloud inferencing or exclude the cloud. Our method balances accelerated edge and cloud resources to maximize task completion and utility. Additionally, unlike prior QoE-based scheduling that assumes static resources or relies solely on cloud scaling, we dynamically adapt to workload fluctuations while minimizing cloud costs.}

\subsection{\addc{Contrast with Our Prior Work}}

\addc{Our earlier work~\cite{10171496}, which this article extends, addressed the QoS metric of maximizing the QoS utility and task completion count within a given deadline using the DEMS and DEMS-A heuristics. These were reported for older DNN models and validated on container-emulated edge hardware. \addcr{In the current version, we have upgraded all the DNN models to recent versions to ensure that our work remains aligned with recent advancements in the field of computer vision.} 
The current article additionally addresses a novel QoE metric for VIP applications that ensures completion of a minimum task count within a window through a QoE utility, and  ensures smooth UAV navigation. These are further integrated as part of an end-to-end VIP navigation application, real edge hardware, and real-world drone experiments reported using contemporary DNN models.}

\begin{figure}[t!]
  \centering
  \includegraphics[width=0.9\columnwidth]{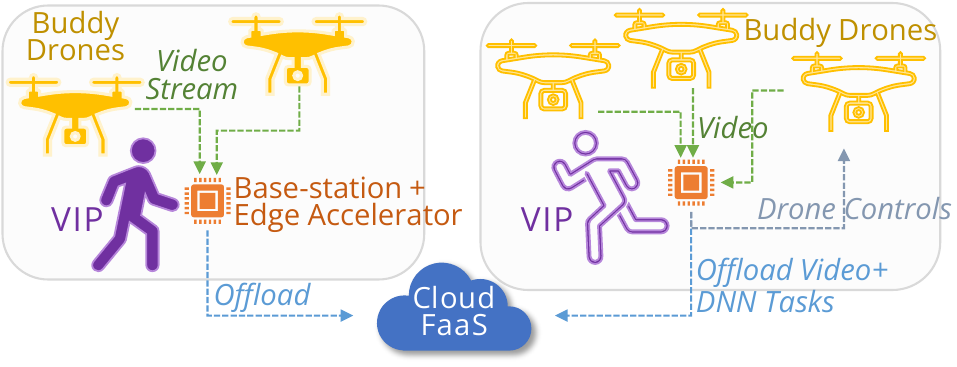}
  \vspace{-0.07in}
    \caption{Buddy Drones Supporting VIP}
    \label{fig:app-model}
    \vspace{-0.2in}
\end{figure}

\vspace{-0.3cm}
\section{A Personalized Drone Platform to Assist VIPs}
\label{sec:vip}
\vspace{-0.2cm}

In this section, we describe our proposed platform architecture for supporting VIP applications while leveraging accelerated edge and cloud. The application and its capabilities are described in more detail in our earlier work, Ocularone~\cite{suman2023chi}.

\subsection{Application}
We assume an environment where a number of Visually Impaired People (VIPs) are using a buddy drone service (Fig.~\ref{fig:app-model}). Each VIP has one or more buddy drones assigned to them with onboard video cameras linked by WiFi to a base station, which is an ARM-based edge computing system with a GPU accelerator. \addcr{These edge accelerators such as Nvidia Jetson Orin Nano are small and weight just $176$~gms, powered using a battery power bank and can fit within the backpack or handbag of the VIP user, without requiring any sophisticated infrastructure.}
Using multiple drones can give better situational awareness of the VIP. The drones may also have varying capabilities, such as the set of onboard sensors, the ability to carry a payload, etc., and these can be used collaboratively.

Our proposed application platform running on the edge base station receives the video streams from the drones, and splits them into a series of short video segments $[ v_1, v_2, \ldots, v_j ]$, each say $1$s or $5$s long depending on the current needs. Multiple ``apps'' may be installed in our platform by the VIP, and each may register a \textit{DNN model} to execute upon the video and consume the response. This result can be used by the app to make some decisions, forming a closed-loop cyber-physical-social system with the buddy drones and the VIP. \modcr{\textit{E.g.}}, an app may detect and warn of incoming traffic, and another may alert if there is a crowd ahead. A navigation app can maintain the VIP in the field of view by controlling the orientation and trajectory of the drone, while another may identify a neighborhood store for pickup and launch a drone to carry a payload. As its features are extended, our platform can serve as an innovative ecosystem to design a variety of personalized drone-based support apps. \modc{We describe the practical use of this app-based platform in the context of the VIP navigation application later in \S~\ref{sec:app-based-framework}.}

\begin{figure}[t]
  \centering
    \includegraphics[width=1.0\columnwidth]{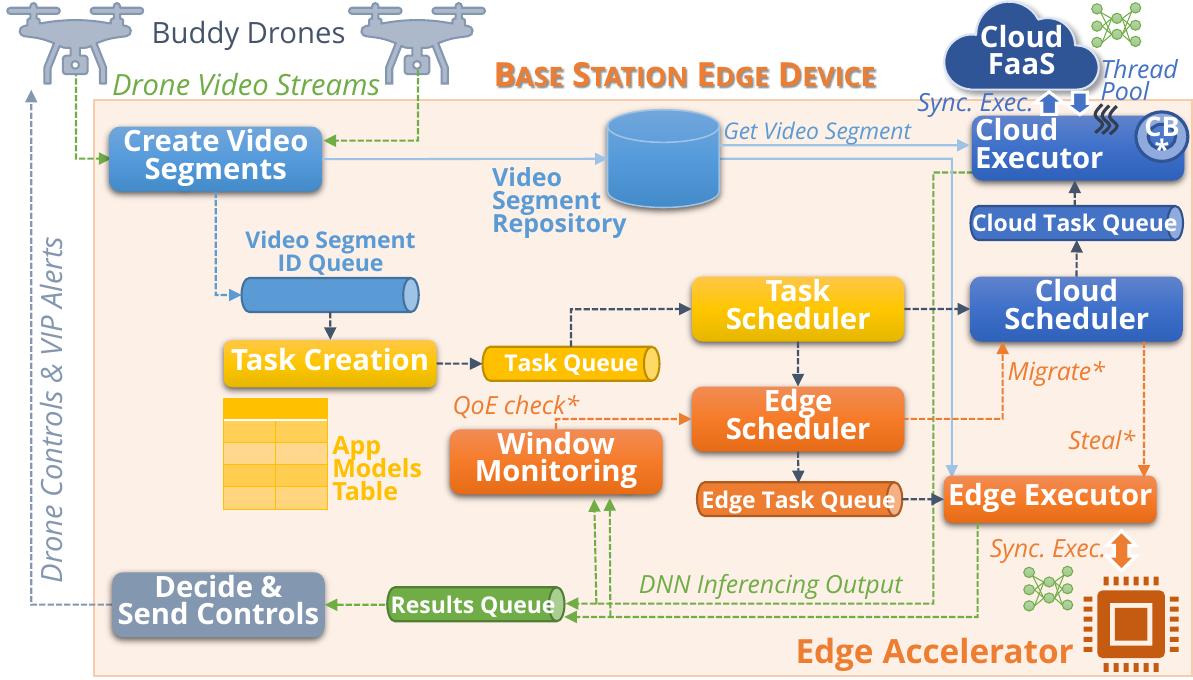}
    \vspace{-0.15in}
    \caption{\modc{Ocularone VIP Application Platform and Inferencing Architecture}}
    \label{fig:arch}
    \vspace{-0.2in}
\end{figure}

As we also discuss later, the apps assign \textit{deadlines} and \textit{benefits} to the models they register, \addcr{which in turn are determined by the apps based on the user requirements, such as ensuring the VIP remains in the field of view of the drone with high confidence, or detecting if the VIP has fallen down within a certain time threshold.}. Given the latency-sensitive nature of the applications, and their impact on the safety of the VIP, such deadlines and benefits help model the temporal importance of the application's tasks and the associated value derived from them. 
\addc{Similary, a user-defined \textit{task completion rate} helps the application to process results from the DNN inferencing output at a regular frequency. 
\modcr{\textit{E.g.}}, DNN inferencing results used to track and follow the VIP must have a higher completion rate to help generate timely control-commands to the drone and ensure a smooth trajectory, while a model that detect a casual point of interest in the vicinity may tolerate a lower rate.}
\delc{\modcr{\textit{E.g.}}, a model to detect incoming traffic may have a higher benefit and a shorter deadline, while one to detect crowds may have a higher benefit but a longer deadline. Models to visually track the VIP can have a high benefit and a very short deadline, but a model to detect a casual point of interest may have a lower benefit and a longer deadline.}

\subsection{\addcr{System Model}}\label{sec:system-model}
\addcr{Our system model involves three key components: the buddy drone, the edge device, and the cloud infrastructure, interconnected via wireless networks. The edge device and cloud infrastructure host services for running DNN models, with the scheduler on the edge device. The edge device is equipped with onboard GPU accelerators, while the cloud FaaS utilizes CPUs to invoke functions that run the models. In our setup, the drone(s) initiate a WiFi hotspot to which the edge device connects. The edge communicates with the cloud over 4G/5G cellular or urban WiFi hotspots, with cloud access over either a Wide Area Network (WAN), which introduces potential bandwidth variability, or a Local/Metropolitan Area Network (LAN/MAN), offering more stable and faster connectivity when cloud resources are local.}


\subsection{Architecture} \label{sec:scheduler-arch}

Fig.~\ref{fig:arch} shows the overall scheduling platform architecture based on a real-world prototype. The buddy drones send continuous video streams to the base station edge device through a WiFi link. A \textit{splitter thread} first segments and stores these video streams as video files of a fixed duration in a video repository maintained in the local file system. An entry for this segment $v_j$ is created and placed in the \textit{video segment FIFO queue}, with the segment ID and the timestamp $t_j'$ at which it was created. The \textit{task creation thread} takes one segment entry $v_j$ at a time from this queue and checks a table of `$n$' \textit{models} that are registered for the VIP at this time by various apps, and creates one inferencing task for each model using this segment ID as input. These $n$ tasks, $[\tau_1^j, \ldots, \tau_n^j]$, are annotated with their deadline durations, $\delta_1 \ldots \delta_n$, \addc{that indicate the required QoS,} and inserted in a randomized order (to avoid favoring any single task) into a task FIFO queue. If $m$ drones are sending video segments to the edge, each would create a set of tasks.

A \textit{task scheduler} thread reads from the task queue and decides whether to pass this task to the \textit{edge scheduler} or the \textit{cloud scheduler}, or \textit{drop} the task. The edge or the cloud scheduler, when it receives the task, puts it in the \textit{edge} or the \textit{cloud task priority queue}, respectively. The decision logic and priorities used are discussed in Sections~\ref{sec:scheduling} \addc{and~\ref{sec:gatekeeper}}.

An \textit{edge executor} takes tasks from the head of the edge task queue and executes them \textit{synchronously} through a \textit{local gRPC DNN inferencing service}. Before that, it does a \textit{Just In Time (JIT)} check to confirm if the deadline is expected to be met if the task is executed on the edge. The gRPC service wraps a PyTorch library that has pre-loaded the active DNN models and executes them on the edge accelerator for the video segment file present in the local video repository. The low-end edge GPUs like Jetson do not have hardware support for concurrent execution of kernels. So a synchronous single-threaded execution ensures a deterministic execution duration for the task on the edge, without contention.

The inference results are returned by the gRPC service to the edge executor, which checks if the deadline for the task execution has not yet expired; if so, it places the tasks's output in the \textit{results FIFO queue}, and selects the next task with the highest priority from the edge task queue to invoke the gRPC service for inferencing. If the deadline for an executed task has expired, its output is stale, and it does not gain the benefit associated with the task even though it is billed for the execution. Apps may choose to use these delayed responses or discard them in favor of more recent videos.

The \textit{cloud executor} is processed by a thread pool, allowing multiple concurrent requests to be sent to the scalable FaaS inferencing. An available thread takes the task at the head of the cloud task queue, performs a JIT check if the expected time on the cloud will return by the deadline, and synchronously executes the FaaS function by passing the video segment file as a \texttt{byte array} over the network. When the function responds, the thread verifies if the deadline is actually met, and puts the output of a successful task in the results queue.

\addc{A \textit{window monitoring thread} keeps a count of the completed tasks per DNN model within a user-specified (tumbling) time window. Based on the current completion rate and the required QoE, the thread identifies tasks from the edge queue and sends it to the cloud (Sec.~\ref{sec:gatekeeper}).}
A \textit{decision thread} analyzes each result from the \textit{results queue}, takes decisions based on the application logic, \modcr{\textit{e.g.}}, change the drone's velocity, or send an ``approaching vehicle'' alert to the user, or do nothing, and notifies the drone or the VIP, if required, through the WiFi link.

Our platform is implemented in Java, with the gRPC services for DNN inferencing implemented in Python and PyTorch. The various queues are implemented using Java concurrent data structures, and we design a custom priority queue for the edge and cloud tasks queues based on a doubly linked list.
The FaaS inferencing is implemented using AWS Lambda that wraps the different DNN models.

\noindent \addcr{The code for the scheduler is available at \url{https://github.com/dream-lab/ocularone-scheduler}.}

\vspace{-0.3cm}
\section{Edge+Cloud Scheduling of Inferencing Tasks}\label{sec:problem}
\vspace{-0.2cm}

Execution of a DNN inferencing model $\mu_i$ on a video segment $v_j$ forms a \textit{task} $\tau_i^j$. The base station edge may execute these tasks locally on the edge accelerator or remotely at a FaaS on the cloud. The relevant DNN models are already deployed in the edge and the cloud, but the video segment needs to be sent to the cloud from the base station. This may be over WAN if the FaaS runs on a public cloud, \modcr{\textit{e.g.}}, AWS Lambda or Azure Functions, or over LAN/WLAN if it is a private cloud, \modcr{\textit{e.g.}}, OpenFaaS in a university campus. The same FaaS deployment is shared by all drones of all VIPs.

An app registering a DNN model $\mu_i$ provides an associated \textit{benefit} $\beta_i$, and a \textit{deadline duration} $\delta_i$ to complete executing its inferencing task for a video segment once it is received at the base station. Tasks that finish execution after the deadline are considered as failed. There is an \textit{estimated execution duration} of $t_i$ for this model's task on the edge accelerator and $\hat{t}_i$ on the FaaS. This is nominally specified based on prior benchmarking for a standard video segment size and selecting, say, the $95^{th}$ percentile of the observed execution times (Fig.~\ref{fig:bm:compute}). We observe this to be stable for the edge accelerator but the FaaS can exhibit variability due to cold-starts~\cite{9191377} and network variability. The cost-per-unit time for executing a task on the edge and the cloud is $\kappa$ and $\hat{\kappa}$. This can be based on the actual billed time for the FaaS, and by amortizing the capital costs and/or using the operational costs for the edge.

A task $\tau_i^j$ may be scheduled on the edge or the cloud for execution. The task may finish within the deadline on the resource, execute but miss the deadline, or be dropped by the scheduler.
Say $\bar{t}_i^j$ or $\hat{\bar{t}}_i^j$ is the \textit{actual execution duration} for a task on the edge or the cloud. 
\modc{The \textit{QoS utility}} $\gamma_i^j$ gained by the task is:
\begin{equation}\label{eqn:base-utility}
\gamma_i^j = 
\begin{cases} 
\beta_i - (t_i \cdot \kappa) & \text{, if task executes on the edge within}\\
                            & \text{~~deadline, $\bar{t}^j_i \leq \delta_i$ }\\
- (t_i \cdot \kappa) & \text{, if task executes on the edge but}\\
                    & \text{~~misses deadline, $\bar{t}^j_i > \delta_i$}\\
\beta_i - (\hat{t}_i \cdot \hat{\kappa}) & \text{, if task executes on the cloud within}\\
                    & \text{~~deadline, $\hat{\bar{t}}^j_i \leq \delta_i$}\\ 
- (\hat{t}_i \cdot \hat{\kappa}) & \text{, if task executes on the cloud}\\
                    & \text{~~but misses deadline, $\hat{\bar{t}}^j_i > \delta_i$}\\ 
0 & \text{, if the task is dropped and not executed}
\end{cases}
\end{equation}

For simplicity, we assume that the cost incurred for a model type is constant for the edge ($t_i \cdot \kappa$) and for the cloud ($\hat{t}_i \cdot \hat{\kappa}$), based on the expected execution time on that resource. But the actual execution duration for a task ($\bar{t}_i^j$ or $\hat{\bar{t}}_i^j$) is used to determine if we truly meet the task's deadline. 

\addc{Further, we introduce a QoE metric that is beneficial to the drone application domain. We have a user-specified \textit{task completion rate}, $\alpha_i$, defined as the fraction of tasks arriving for a DNN model $\mu_i$ within a tumbling time window $W_i$ of duration $\omega_i$ that must complete within their deadline $\delta_i$. The window is initially formed at the start of the application and all tasks having a completion time within the next $\omega_i$ seconds are part of this window. The next time window starts at the end of the previous window. To generalize this across application scenarios, we allow $\alpha_i$ and $\omega_i$ to be specific to each DNN model $\mu_i$.}

\modcr{If tasks arriving for a DNN model meet the completion rate in a time window, we assign an additional benefit $\bar{\beta_i}$ to the window, and calculate \textit{QoE utility $\bar{\gamma_i}$} as:}
\begin{equation}\label{eqn:bonus-utility}
\bar{\gamma_i} = 
\begin{cases}
        \bar{\beta_i} &  \text{, if at least $\alpha_i$ fraction of tasks finishing within $W_i$} \\
        & \text{~~for DNN model $\mu_i$ complete within their deadline} \\
        0 & \text{, otherwise}
\end{cases}
\end{equation}
\addc{Our \textit{total utility} is given by the sum of QoS utility for all tasks and the QoE utility for all windows, for all DNN models: $\gamma = \sum_j\sum_i\gamma_i^j + \sum_W\sum_i\bar{\gamma_i}$.}
\\
\textbf{\bf Optimization Problem.}
\textit{For a stream of tasks $\tau_i^j$ created at the base station from one or more drones, schedule them on edge or cloud resources \modc{such that: (i) the QoS utility is maximized,} (ii) the number of tasks completed within the deadline is maximized, \addc{and (iii) the QoE utility is maximized.}} 

\modc{We solve this in two parts. Initially, we treat the first two objectives of maximizing QoS utility and successful task count as a dual-objective problem which we individually maximize using \DS.}
This maximization is considered both from the perspective of a single VIP having one or more drones linked to their base stations, as well as for all the VIPs supported by the buddy drone deployment. \addc{Next, we maximize the total utility $\gamma$, which includes the third objective, using \GS, which extends \DS and tries to guarantee a better QoE.} 

\vspace{-0.3cm}
\section{Deadline-aware Scheduling Heuristics} \label{sec:scheduling}
\vspace{-0.2cm}

In this section, we propose a scheduling algorithm \textbf{\DS} to solve the optimization problem across the edge and cloud resources. Specifically, this scheduler for the buddy \uline{\textbf{D}}rones combines several heuristics -- \uline{\textbf{E}}arliest deadline first priority ordering on the edge queue, task \uline{\textbf{M}}igration from edge to cloud queue to avoid deadline misses on edge, work \uline{\textbf{S}}tealing from cloud to edge queue to leverage slack on the edge, and an \uline{\textbf{A}}daptive variant (\textbf{\DS-A}) that responds to wide area network variability. These are described below.

\subsection{\uline{E}arliest Deadline First Priority}\label{sec:edf}
The task scheduler, by default, attempts to send an input task to the edge scheduler. The \textit{edge scheduler} initially starts with the \emph{Earliest Deadline First (EDF)} scheduling algorithm to assign the \textit{task priority} for its edge task queue. In EDF, tasks with shorter deadlines get a higher priority, and tasks with longer deadlines are pushed to the rear. The priority for a task $\tau_i^j$ is given by $(t_j' + \delta_i)$, where $t_j'$ is the time at which the video segment is created at the base station. 

When a task is placed in its priority position within the queue, we do a \textit{feasibility check} on whether the sum of the execution times of tasks ahead of this new task, plus its own expected execution time, exceeds its deadline; if so, the edge scheduler does not accept this task and instead is directed to the cloud scheduler.

The \textit{cloud scheduler} uses a FIFO cloud task queue. Before accepting a task from the task scheduler, it will check if: (1) the task can complete within its \textit{deadline} if scheduled immediately on the cloud, and (2) the utility expected from this task executing on the cloud is \textit{non-negative}. Since the cloud executor uses a thread pool to invoke the FaaS, the cloud task queue will be drained rapidly as its task requests are sent before responses from previous tasks have arrived.

Starting from this baseline deadline-driven scheduling logic across edge and cloud, which we call \textbf{\EC}, we enhance logic with several additional heuristics. Existing work~\cite{8917749,dai2019scheduling,7300228} have also used a variation of such baselines for scheduling latency-driven applications. 

\subsection{\uline{M}igrating Tasks from Edge to Cloud Queue} \label{sec:migration}

By default, the edge scheduler only checks if an incoming task will violate its own deadline when it is inserted into the relevant priority position in the \EDFe edge queue. However, as a result of inserting a task ahead of other lower priority tasks already in the queue, these lower priority tasks may no longer be able to meet their deadlines. So, the \EC baseline only offers a new task to the cloud scheduler if it is expected to miss its own deadline due to higher priority tasks ahead of it in the edge queue. It is possible that tasks that appear to miss their deadlines at the rear of the queue may actually complete on time, \modcr{\textit{e.g.}}, if the tasks ahead of it finish executing on the edge faster than their expected execution duration. But there is a risk of the deadline miss on the edge.

We introduce a \textit{migration heuristic} to check if the existing tasks behind a new task being inserted into the edge priority queue will now miss their deadlines. For this, we introduce $S_i^j$ as the \textit{score} of a task which is calculated as: 
\begin{equation}
S_i^j = 
\begin{cases} 
\gamma_i^E - \gamma_i^C & \text{, if task executes on the cloud within}\\
                            & \text{~~deadline and $\gamma_i^C > 0$ }\\
\gamma_i^E  & \text{, if task executes on the cloud but}\\
                    & \text{~~misses deadline or $\gamma_i^C \leq 0$}
\end{cases}
\end{equation}

\noindent
\addcrr{where $\gamma^E$ and $\gamma^C$ are the utility values assigned to successful tasks executed on edge and cloud respectively, based on Eqn.~\ref{eqn:base-utility}.}
We calculate the scores of all lower priority tasks that will miss their deadlines in the edge queue due to insertion of the new task. If none miss their deadline, the new task is inserted into the edge queue. Else, if the sum of scores of edge tasks missing their deadline is less than the score of the incoming task, we \textit{migrate} these existing edge task(s) to the cloud queue by offering them to the cloud scheduler. But, if the score of the new task is lower, we retain existing tasks in the edge queue and instead redirect the incoming task to the cloud queue. Any task not accepted by the cloud scheduler, \modcr{\textit{e.g.}}, due to deadline miss or negative utility on the cloud, is dropped.

\begin{figure}[t]
  \centering
    \includegraphics[width=1.0\columnwidth]{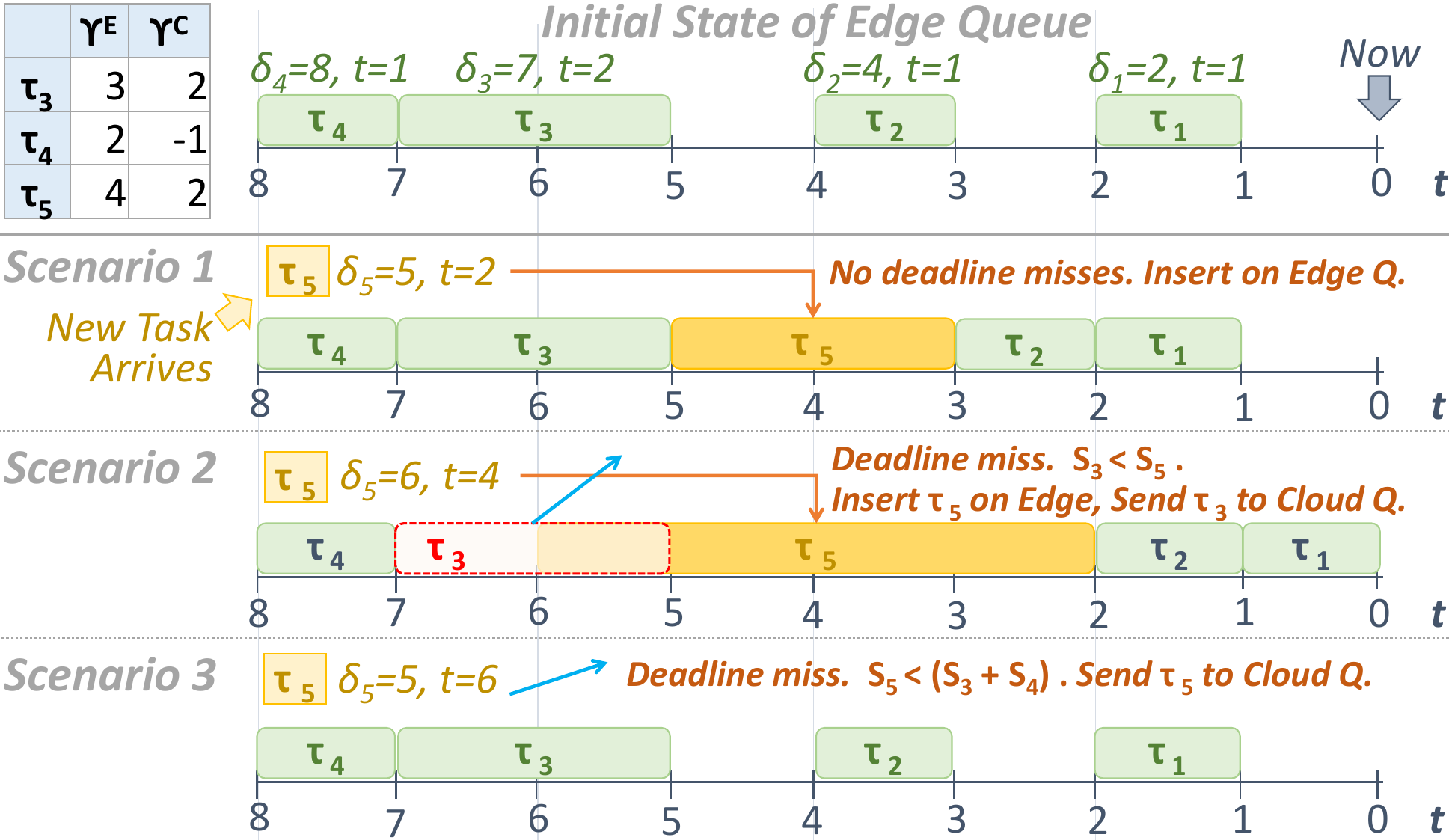}
    \vspace{-0.1in}
    \caption{Migrating Tasks from Edge to Cloud Queue
    }
    \label{fig:migration}
    \vspace{-0.2in}
\end{figure}

We illustrate these three migration scenarios in Fig.~\ref{fig:migration}. The initial state of the edge queue is shown at the top, with the tasks $\tau_i$ in EDF order and their deadlines $\delta_i$ and expected edge execution duration $t_i$ shown. In \textit{Scenario 1}, an incoming task $\tau_5$ does not cause any deadline -- of itself or an existing edge task -- to violate. In \textit{Scenario 2}, $\tau_5$ will lead to deadline violation of $\tau_3$. \modcr{We calculate the scores of $S_3 = 1$ and $S_5 = 2$, and since $S_3$ is smaller}, we migrate $\tau_3$ to the cloud queue and insert $\tau_5$ in the edge queue. In \textit{Scenario 3}, insertion of $\tau_5$ will cause deadlines  of $\tau_3$ and $\tau_4$ to be missed. The scores are $(S_3 + S_4 = 3) > (S_5 = 2)$. So, $\tau_3$ and $\tau_4$ are retained in the edge queue, and $\tau_5$ is offered to the cloud queue. 

\subsection{Work \uline{S}tealing from Cloud to Edge Queue} 

By sending tasks to the cloud, we may increase the chances of the task being completed. But given that the FaaS is usually costlier than the edge, the utility gained for such tasks may be less. In fact, by sending more tasks to the cloud if they could have instead been executed on the edge, we are reducing the overall utility in favor of task completion count. Tasks on the edge may execute earlier than expected -- due to transient over-performance of the edge or over-estimation of the edge execution times for some task types. This opens up slack on the edge, which we exploit using a \textit{work stealing heuristic}.

Here, when the edge executor is ready to execute the next task at current time $t''$, it \textit{peeks} into the edge task queue and checks the deadline of the head task $\tau_i^j$ against its expected execution time. Say the \textit{slack time} available for this task is $\sigma_j = (t'_j + \delta_i) - (t'' + t_i)$. If $\sigma_j > 0$, then the task is expected to complete on the edge before its deadline, if executed now. If $\sigma_j$ is greater than the smallest model execution time on the edge, $\min_k(t_k)$, it may be possible to ``steal'' a task from the cloud queue to be executed immediately on the edge, provided: (i) the stolen task is expected to complete on the edge within its deadline, and (ii) executing the stolen task on the edge will not cause the deadlines of existing tasks on the edge queue to be violated. We can also incrementally steal multiple tasks based on the available slack.

Tasks that can be stolen from the cloud queue are the ones that can fit within the available slack on the edge. Among these eligible tasks, we calculate a \textit{rank}, given by $\frac{\gamma_i^E - \gamma_i^C}{t_i}$, \modcr{\textit{i.e.}}, the difference between the expected utility from executing the task on the edge ($\gamma_i^E$) and the expected utility from executing it on the cloud ($\gamma_i^C$), divided by the expected execution duration on the edge ($t_i$). Intuitively, this prefers tasks that will give a better utility gain per unit execution time on the edge. Further, among these, we prioritize tasks with a \textit{negative cloud utility} since they will otherwise be dropped by the cloud executor. 

We make a few enhancements to the default scheduling design (Sec.~\ref{sec:scheduler-arch}) to leverage this strategy. Instead of the cloud executor immediately sending tasks in the cloud task queue for execution on the FaaS, we instead assign it a \textit{trigger time}, which is the difference between the deadline for the task and its expected execution duration on the cloud, plus a safety margin. The cloud task queue uses this \textit{trigger time as its priority} rather than the default FIFO. The cloud executor sends a task from the cloud \textit{priority} queue to the FaaS for execution only when the current time reaches its trigger time. This allows tasks intended for the cloud to be \textit{deferred for execution}, giving them a chance to be stolen by the edge if there is slack, and yet be executed on the cloud on time if not stolen.

\begin{figure}[t]
  \centering
    \includegraphics[width=1.0\columnwidth]{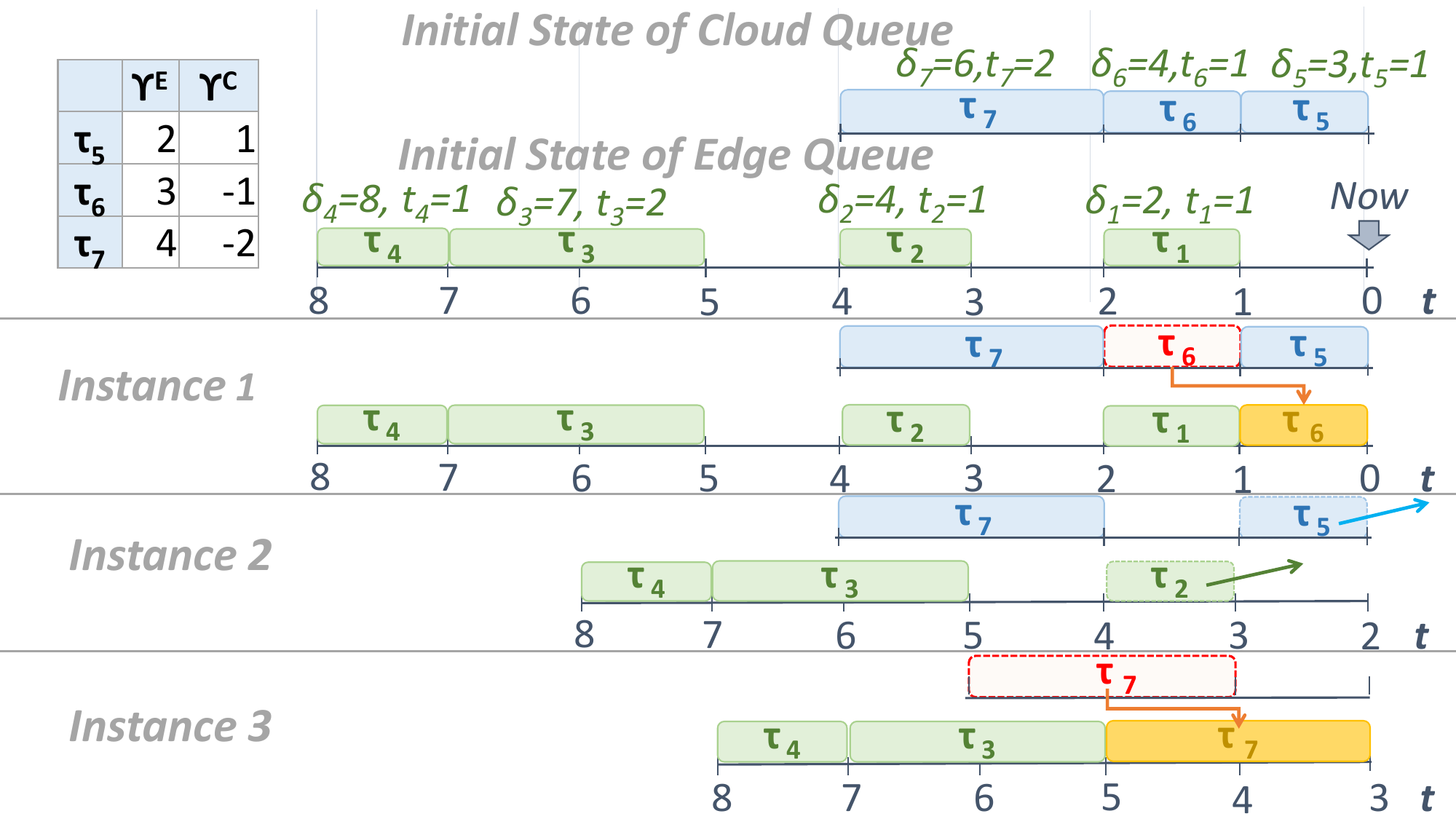}
    \vspace{-0.1in}
    \caption{Stealing Tasks from Cloud to Edge Queue}
    \label{fig:stealing}
    \vspace{-0.2in}
\end{figure}

The cloud scheduler also allows tasks with \textit{negative utility} on the cloud to be placed on the cloud queue, rather than drop them immediately --
this gives them a chance to be stolen by the edge while waiting in the queue for their trigger time. Here, we set its trigger time to the latest time the task needs to be executed \textit{on the edge}. Since the edge trigger time is typically later than the cloud trigger time, this approach gives more time for the task to be stolen. If a negative utility task hits its trigger time and has not been stolen, the cloud executor drops it JIT.

We illustrate work stealing in Fig.~\ref{fig:stealing}, where the initial state of the edge and cloud queue are shown. For simplicity, we do not account how the tasks have been inserted in the cloud queue and instead assume that tasks with specified properties are present in the cloud queue with the assumption that $t_i = \hat{t_i}$. At instance 1, slack duration of $1 unit$ is available, for which $\tau_5$ and $\tau_6$ are eligible, but since we prioritize the tasks with a negative utility on cloud, $\tau_6$ is stolen and executed on the edge with subsequent execution of $\tau_1$. At instance 2, $\tau_5$ and $\tau_2$ get triggered for cloud and edge executions, respectively, based on their trigger times. Instance 3 computes a slack of 2 units which pulls $\tau_7$ from the cloud queue and results in a better edge utilization.

\subsection{\uline{A}dapting to Variability} 

One challenge with executing tasks on public clouds is the network variability over the WAN or, in the case of private clouds, any capacity constraints that cause queuing of inferencing requests. VIPs moving in urban areas can encounter high variability in 3G/4G bandwidth (Fig.~\ref{fig:bm:nw:wan-bw}). Both these factors may be transient.

By default, the cloud scheduler and executor determine if a task can be executed on the cloud within its deadline based on a \textit{static} expected execution duration provided for a model from prior benchmarks. To account for the above variability, the cloud executor maintains a \textit{sliding window average ($\bar{\bar{t_i}}$)} of the actual observed execution duration $\bar{t}_i^k$ for inferencing by each DNN model $\mu_i$ on the cloud, using a circular buffer of size $w$ (CB, in Fig.~\ref{fig:arch}). When this observed average time, $\bar{\bar{t_i}} = \frac{\sum_{k\in w}{\hat{\bar{t}}_i^k}}{w}$, exceeds the expected time $\hat{t}_i$ by a threshold, $\bar{\bar{t_i}} - \hat{t}_i > \epsilon$, we update the expected execution duration for this model on the cloud to be, $\hat{t}_i = \bar{\bar{t_i}}$. When this happens, we use this expected execution duration to compute the trigger time and feasibility of future tasks of this model that arrive at the cloud queue.

It is also possible that there is a period of high execution latency, and the updated execution duration becomes so high that none of the future tasks for the DNN model will complete within their deadline. In such cases, no future tasks of this model type will be sent to the cloud, which means the execution duration will never be updated even if the latency improves after some time. To avoid such a point of no return, we set a maximum \textit{cooling period ($t_{cp}$)} during which tasks are skipped due to expected deadline misses. After this period elapses, we reset the expected execution duration for the model to its static default and attempt to discover its current execution duration on the cloud from new tasks that are sent to the cloud.
We set $w=10$, $\epsilon=10$~ms, and $t_{cp}=10$~s in our experiments based on empirical observations.

\vspace{-0.3cm}
\section{\addc{Guaranteeing Quality of Experience (QoE) to Users}}\label{sec:gatekeeper}
\vspace{-0.2cm}

\addc{In this section, we build upon \DS to propose the \GS algorithm, which employs an adaptive heuristic to guarantee the QoE goal specified by the user.
\GS help achieve the required minimum fraction of task completion within their deadline in a given time window. Intuitively, it keeps track of the task completion rate for each model for its current window, which may fall below the threshold due to delays in the edge queue or edge inferencing duration.
It responds by potentially moving pending tasks from the edge to the cloud queue to improve the completion rates before the end of the window such that the target completion rate is eventually met and the QoE utility is accrued for the window.
\GS relies on the cloud to enhance task completion rates and QoE utility, but this is possible only for tasks with a non-negative utility on the cloud.
}

\begin{algorithm}[t]
\caption{\addc{GEMS triggered after each completed task}}\label{algo:gatekeeper}
\small
\begin{algorithmic}[1]
\Function{GEMS}{$\mu_i, \alpha_i, x, isSucc, w^s_i, w^e_i$}\\
f-    \Comment{\emph{$x \in (w_i^s,w_i^e]$ is current timestamp for completion of a task of $\mu_i$ falling within current window, with $\alpha_i$ being the minimum completion rate. $isSucc$ is \texttt{True} if the task completed within its deadline.}}
    \State $\lambda_i = \lambda_i + 1$ \Comment{\emph{Update total task count for $\mu_i$}}
    \If {$isSucc==$\texttt{True}} \label{algo:is-success}
    \State $\widehat{\lambda_i} = \widehat{\lambda_i} + 1$ \Comment{\emph{Update successful count for $\mu_i$}} \label{algo:success-incr}
    \EndIf
    \State $\widehat{\alpha}_i =\frac{\widehat{\lambda_i}}{\lambda_i}$ \Comment{\emph{Update incr. rate for $\mu_i$ in this window}}\label{algo:line7}
    \If{$\widehat{\alpha}_i$ < $\alpha_i$} \Comment{\emph{If we fall behind...}}\label{algo:line8}
    \For {$\tau_i^j \in \mu_i$ in edge queue}  \label{algo:line10} \\
    \Comment{\emph{Consider migrating pending edge tasks to cloud.}}
    \If {$(\gamma_i^C > 0) ~~ \&\& ~~ (x + \hat{t_i} \leq t_j' + \delta_i) $ }
    \State Send $\tau_i^j$ from edge queue to the cloud queue 
    \EndIf
    \EndFor  \label{algo:line14}
    \EndIf  \label{algo:line15}
    \If {$x==w^e_i$} \Comment{\emph{If we reach end of current window...}}
        \If{$\widehat{\alpha}_i \geq \alpha_i$} \Comment{\emph{If window rate meets threshold...}} \label{algo:qoe-check}
            \State $\Bar{\Bar{\gamma_i}} ~~+= \bar{\gamma_i}$ \Comment{\emph{Update QoE utility}} \label{algo:qoe-incr}
        \EndIf
        \State $w^s_i = w^e_i$, $w^e_i = w^e_i + \omega_i$ \Comment{\emph{Increment window.}} \label{algo:line-reset1}
        \State $\lambda_i = 0$, $\widehat{\lambda_i}=0$, $\widehat{\alpha}_i=0$  \Comment{\emph{Reset counts and rates.}} \label{algo:line-reset2}
    \EndIf
\EndFunction
\end{algorithmic}
\end{algorithm}

\subsection{\addc{Strategy}}
\addc{
The user provides a list of \textit{required task completion rates}, $\alpha = [ \alpha_1,...,\alpha_n ]$ to be achieved within each \textit{tumbling time window duration}~\footnote{\addc{We use tumbling windows for their simplicity of specification by the users. \textit{Tumbling windows} are non-overlapping and partition the wallclock time into windows of the specified width, unlike \textit{sliding windows} which can overlap based on the sliding factor.}}, specified by $[\omega_1,..., \omega_n]$, for $n$ DNN models $[\mu_1,...,\mu_n ]$, with $\alpha_i > 0$. The associated QoE utility for each successful window is represented by ${\Bar{\gamma}}= [ {\Bar{\gamma}}_1,...,{\Bar{\gamma}}_n]$, while the overall QoE utility is given by $\Bar{\Bar{\gamma}}$.
The \textit{window monitoring thread} (Algorithm~\ref{algo:gatekeeper}) runs the GEMS algorithm to continuously calculate the \textit{incremental task completion rate} $\widehat{\alpha}_i$ within the current window for every model $\mu_i$ after each task is processed. It checks if the completion rate is falling below the required rate, $\widehat{\alpha}_i < \alpha_i$, and if so, uses \GS 
to try and improve the rate. This is done by greedily moving all pending tasks of the lagging model from the edge queue to the cloud queue, provided we get a positive utility on the cloud and can finish within the deadline for each task.}

\addc{Specifically, we define a window $W_i=(w_i^s,w_i^e]$ for $\mu_i$, which is initially formed at the start of the application and subsequently ``tumbles'',
with $w_i^s$ and $w_i^e$ being the start and end timestamp of the window, \modcr{\textit{i.e.}}, $w_i^e = w_i^s + \omega_i$. Now, at any timestamp $x$ within the window, $w_i^s \leq x < w_i^e$, we define the incremental task completion rate as $\widehat{\alpha}^x_i =\frac{\widehat{\lambda^x_i}}{\lambda^x_i}$. This is the number of tasks for model $\mu_i$ whose completion time falls within the window and they successfully finish within their deadline ($\widehat{\lambda^x_i}$), out of all the tasks of $\mu_i$ in the window that have finished until time $x$ ($\lambda^x_i$).
}

\addc{For each task for model $\mu_i$ that is completed (or dropped) at some time $x \in (w_i^s,w_i^e]$ by the \textit{edge or cloud executor thread}, we trigger Alg.~\ref{algo:gatekeeper}. It performs the incremental check on whether the task's deadline was met relative to its arrival time, $x \leq t_j' + \delta_i$, and only then increments the successful tasks count $\widehat{\lambda^x_i}$ by $1$ for the $x^{th}$ timestamp within the window (lines~\ref{algo:is-success}--\ref{algo:success-incr}).}

\addc{
It then updates the incremental completion rate $\widehat{\alpha}^x_i$ (line~\ref{algo:line7}) and compares it with the required task completion rate $\alpha_i$ (line~\ref{algo:line8}). If we are falling behind, \modcr{\textit{i.e.}} $\widehat{\alpha}^x_i < \alpha_i$, we scan the edge queue starting from the head for all tasks of $\mu_i$ waiting to execute on the edge.
If the task (1) has a non-negative utility if executed on the cloud ($\gamma^C_i>0$), and (2) can be completed within its deadline if sent to the cloud immediately ($x + \hat{t_i} \leq t_j' + \delta_i$), then it is immediately sent to the cloud (lines~\ref{algo:line10}--\ref{algo:line14}).
This repeats for each task successively until all the tasks of model $\mu_i$ in the edge queue have been considered. This scan happens on a task completion trigger only if the incremental completion rate falls below the required rate.}

\addc{Finally, when we reach the end of the current window, if the final task completion rate meets the required QoE rate, \GS adds the QoE utility accrued by the model $\mu_i$ for that window, $\bar{\gamma_i}$, using Eqn.~\ref{eqn:bonus-utility} to our overall utility, $\Bar{\Bar{\gamma_i}}$~(lines~\ref{algo:qoe-check}--\ref{algo:qoe-incr}). 
We then increment to the new time window, reset the counters and rates for it (lines~\ref{algo:line-reset1}--\ref{algo:line-reset2}), and repeat this process.
In our experiments, we set $\omega_i = 20s$ for all models for simplicity and vary $\alpha_i$.}

\ysnoted{Three major problems with this approach. One, the trigger time is when the task completes. Which means that we may be stuck without any response for tasks for a model that are getting delayed for long and not be able to respons. Two, we are doing this check based on the completion timestamp rather than arrival timestamp for the tasks, which is an error. Lastly, we send \textit{all pending tasks} in the edge queue to the cloud rather than sufficient ones to meet the QoE utility. It is inefficient. Need to live with it for now.}

\vspace{-0.3cm}
\section{\addc{Integrated and Responsive Analytics for VIP Navigation}}\label{sec:app-based-framework}
\vspace{-0.2cm}
\begin{figure}[t]
  \centering
    \includegraphics[width=0.8\columnwidth]{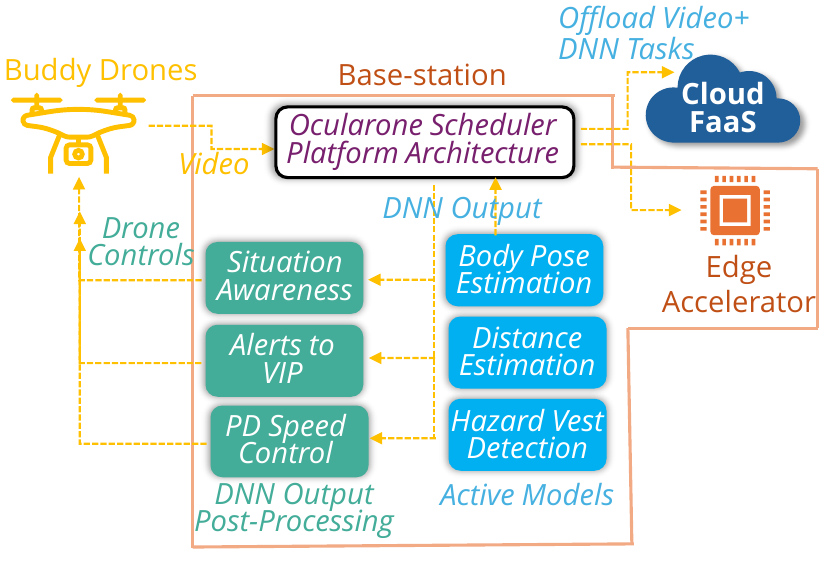}
    \vspace{-0.07in}
    \caption{\addc{Integrated analytics framework for VIP Navigation showing post-processing of the output of different DNN models}}
    \label{fig:app-based-framework}
    \vspace{-0.2in}
\end{figure}

\addc{In this section, we examine the use of our scheduling architecture and heuristics to support the Ocularone VIP navigation application, as illustrated in Figure~\ref{fig:app-based-framework}. The application uses several DNN models, as explained next. Based on the VIP's preferences and current conditions, the application registers these as active DNN models with the scheduler, along with their QoS deadlines, the QoE completion window rates and durations, and their utilities. The drone(s) deployed to assist the VIP generate a stream of video frames that are processed by our scheduler and their output returned to the application, which then takes any action that is required by interfacing with the drone(s), the VIP and/or external services.}

\addc{One of these critical activities, which we also validate experimentally in \S~\ref{subsec:hardware-validation}, is for the buddy drones to identify and autonomously follow the VIP they are assigned to. The VIP wears a hazard vest as an identification marker. The drone detects the marker by running a specialized \textit{Hazard Vest (\HV) detection} DNN model, based on a retrained Yolov8 nano model~\cite{yolov8_ultralytics}, over the video frames from its front-facing camera. This ensures that the vest remains within the FoV of the camera as the VIP moves in different directions and at diverse speeds.
Each \HV task executed on a video frame returns a bounding box of the hazard vest worn by the VIP, which is then used by our \textit{Proportional Derivative (PD) Control loop algorithm}~\cite{nagrath2006control} to determine the offset of the vest from the center of the camera frame. This is used to generate a series of control signals to alter the drone's direction of motion along its degrees of freedom: up, down, forward, backward, clockwise or anti-clockwise rotation (yaw), and the speed in that direction. This ensures that the drone maintains the necessary orientation to the VIP -- facing their rear and with a view in front of them -- and proximity under dynamic conditions~\cite{suman2023chi}. The QoS and QoE parameters are set by the application to ensure that the navigation is robust, based on whether the VIP is walking, running, turning, etc.}

\addc{There are other assistive activities that are supported by our VIP application. One of them identifies scenarios when the VIP may require assistance but it unable to explicitly ask for it, \modcr{\textit{e.g.}}, if they fall down or sit down due to exhaustion.
We detect such conditions using a \textit{Body Pose (\BP) estimation} DNN model, based on ResNet18~\cite{resnet18}, which is configured to identify specific poses of concern. Running a \BP inferencing task by the scheduler on the cropped video frame containing the VIP returns the coordinates of 18 body landmarks. This is post-processed by a Support Vector Machine (SVM) based classifier~\cite{vapnik1999nature} that returns one of these categories: upright, kneel, fall, start/stop, land. These are configured to trigger various situation awareness actions, \modcr{\textit{e.g.}}, a fallen pose can trigger the drone to reduce its altitude, get close to the VIP and notify an emergency contact with their photo. Here too the QoS and QoE parameters are tuned for this use case.}

\addc{Other models used by the application to \textit{estimate the distance} from the drone to the VIP, such as \DEV based on Yolo v8 nano~\cite{yolov8_ultralytics} coupled with a linear regression model that post-processes the height, width and area of the VIP bounding box to estimate the absolute distance. 
Another estimates the distances between the VIP/buddy drone and nearby obstacles using \DEO, based on Monodepth2 monocular depth-map model~\cite{monodepth2github}.
These leverage vision and geometric strategies to alert the VIP to ambient condition, and to help the VIP and/or drone navigate around them to avoid a collision. Here again, QoS and QoE are configured by the application for the type of environment, \modcr{\textit{e.g.}}, urban streets with fast-moving vehicles may require a faster and more sensitive responses compared to walking in a park. Lastly, \textit{Mask Detection (\MD)} based on the SSD model~\cite{ssd} and \textit{Crowd Density (\CD) estimation} using Yolov8 (medium)~\cite{yolov8_ultralytics} provide ambient information on their surroundings to the VIP, alert them if nearby individuals are not wearing masks during COVID, and make them feel more inclusive.}

\vspace{-0.3cm}
\section{Experiments} \label{sec:results}
\vspace{-0.2cm}

\addc{We first evaluate the \GS heuristic along with other baselines on improving the QoS utility, on-time completion of task, adaptation to network variability and scaling (\S~\ref{sec:exp:baselines}--\ref{sec:exp:scaling}). We use multiple workloads for multiple drones, executed using emulated Jetson Nano edge accelerators and AWS Lambda cloud functions (\S~\ref{sec:exp:setup}). We extend this to evaluate the QoE utility improvements for \GS using alternative edge and cloud configurations (\S~\ref{subsec:qoe-results}). We lastly perform field experiments using real drone hardware, Jetson Orin Nano accelerator and AWS Lambda functions to validate a VIP application (\S~\ref{subsec:hardware-validation}).}

\subsection{Setup}\label{sec:exp:setup}
\modc{We evaluate the \textit{six DNN inferencing models} described above for VIP drone applications: Hazard Vest detection (\HV)~\cite{raj2025ocularonebenchbenchmarkingdnnmodels}, Body Pose estimation (\BP)~\cite{resnet18}, Distance Estimation to VIP (\DEV)~\cite{raj2025distanceestimationsupportassistive}, Distance Estimation to Objects (\DEO)~\cite{raj2025distanceestimationsupportassistive}, Crowd Density estimation (\CD)~\cite{yolov8_ultralytics} and Face-Mask Detection (\MD)~\cite{ssd}.}
The characteristics of these models are given in Table~\ref{setup-table}, based on real-world benchmarking, \addc{system characteristics and application requirements}. Here, $\beta$ is a relative normalized benefit of the model with no units, while $\delta, t$ and $\hat{t}$ are task deadlines and edge and cloud latencies in $ms$. \addc{These are determined by the VIP application.} \modcr{$\mathcal{K}$} and \modcr{$\hat{\mathcal{K}}$} are the normalized monetary costs on the edge and the cloud. \addc{These are based on real-world amortized costs for performing inferencing on the Jetson Nano edge accelerator~\cite{orin-tech-specs} and executing FaaS inferencing on AWS for different models, as described in \ref{appendix:costing}.}
\modc{For the expected inference latencies on edge ($t$) and cloud ($\hat{t}$),} we use the average of the $95^{th}$ percentile latency for the FaaS invocations measured with concurrent clients, and average of the $99^{th}$ percentile for the edge gRPC invocation using concurrent clients, as described in \ref{appendix:benchmarking}. Note that \BP, shaded in pink in Table~\ref{setup-table}, has negative utility when executed on the cloud, \modcr{\textit{i.e.} $\beta < \hat{\mathcal{K}}$.}

\begin{table}[h]
\centering
\caption{\modcr{Workload configs. of DNN tasks on Jetson Nano to evaluate \DS.} \tablefootnote{\addcr{$\gamma^E$ and $\gamma^C$ represent the values assigned to successful tasks based on Eqn. 1. Tasks that miss their deadline are assigned $\mathcal{-K}$ or $\hat{\mathcal{-K}}$, depending on whether they were executed on the edge or cloud, or 0 if not executed at all.}}}
\label{setup-table}
\vspace{-0.07in}
\footnotesize
\setlength{\tabcolsep}{4.5pt}
\begin{tabular}{l||rrrrrrrr||@{}c@{}|@{}c@{}}
\toprule
\textbf{DNN} & $\beta$ & $\delta$ & $t$ & $\hat{t}$ & $\mathcal{K}$ & $\hat{\mathcal{K}}$ &  $\gamma^E$ & $\gamma^C$ &
\textbf{P}assive & \textbf{A}ctive\\
\midrule
\HV & 125 & 650 & 174 & 398  & 1 & 25 & 124 & 100 & \checkmark & \checkmark\\ 
\hline
\DEV & 100 & 750 & 172 & 429  & 1 & 26 & 99 & 74 & \checkmark & \checkmark \\
\hline
\MD & 75 & 850 & 142 & 589 & 1 & 15 & 74 & 50 & \checkmark & \checkmark \\  \hline
\cellcolor{pink}\BP & \cellcolor{pink}40 & 900 & 244 & 542 & 2 & \cellcolor{pink}43 & 38 & \cellcolor{pink}-3   & \checkmark & \checkmark \\ 
\hline
\CD & 175 & 1000 & 563 & 878  & 4 & 152 & 171 & 23 & $\times$ & \checkmark\\\hline
\DEO & 250 & 950 & 739 & 832 & 6 & 210 & 244 & 40  & $\times$ & \checkmark\\ 
\bottomrule 
\end{tabular}
\vspace{-0.1in}
\end{table}

We create \textit{two application workloads} for the VIPs which have a mix of these models \addc{to evaluate the QoS utility}: \textbf{A}ctive (all six \addc{tasks}) and \textbf{P}assive (\HV, \MD, \modcr{\DEV}, \BP). \addcr{Passive mode suits slow-moving or sparse environments, while Active mode is designed for more busy scenarios.}
These are combined with $2$, $3$ or $4$ buddy drones (2D/3D/4D) per VIP that concurrently send video streams to an edge base station for the VIP. This gives us 6 possible workloads which generate between $8$--$24$ tasks/second per edge. In our experiments, each run generates between $16k$--$201k$ tasks depending on the workload and number of drones.

We use the \textit{VIoLET} Internet of Things (IoT) emulation environment~\cite{badiger2018violet} to launch containers that map to an \textit{Nvidia Jetson Nano} edge base station, and containers that model \textit{Ryze Tello drones} connected to the edges over a network link that emulates the performance of WiFi. The containers run on a host server having an Intel Xeon 6208U CPU@$2.9$~GHz with $30$~vCPUs and $128~GB$ RAM, running \textit{Ubuntu v18.04}.
Each edge container is allocated $4$~vCPUs and $4$~GB RAM to match the inferencing performance of the Nano, and a drone container is lightweight with $0.1$~vCPU and $100$~MB RAM. 

We collect real video feeds on campus using the Tello drones and stream these from the drone containers to their base station containers. Our software platform, scheduler logic, and gRPC service hosting the DNN models run on each edge container. Our platform creates $1$s segments from these video streams, each $\approx 38$~kB. One edge container serves $1$--$4$ drones, based on the workload. A single host server has $7$ edge and $7$--$28$ drone containers and connects from the Indian Institute of Science university campus in Bangalore, India to FaaS running on the AWS data center at Mumbai, India. \modc{Each Lambda function is allocated $\{ 2,2,1,4,2,5 \}$~GB of RAM for the $\{ \HV,\DEV, \MD,\CD, \\ \BP,\DEO \}$ models, respectively, to meet the end-to-end execution deadlines for each model.} The normalized edge and cloud costs in Table~\ref{setup-table} are based on real costs paid for running the DNN tasks on the edge (amortized) and AWS Lambda.

\subsection{Baseline Scheduling Algorithms}
\label{sec:exp:baselines-desc}

We compare \textit{\DS} against several baseline scheduling algorithms, both classic and contemporary, on the QoS utility and task completion count. \modcr{We consider cloud-only and edge-only strategies proposed by us, edge + cloud strategies adapted and enhanced from classical scheduling algorithms, and state-of-the-art (SOTA) strategies from \cite{8941266, fu2022kalmia, 10.1145/3492321.3519576} as baselines.} \delcr{\emph{Shortest} \textit{Job First (\SJFe)} prioritizes tasks with the least expected execution duration on the edge queue ($t_i$). This should complete more short tasks on the edge, and achieve high task throughput but with reduced utility and possible deadline misses.} \delcr{\textit{Highest} \textit{Utility First (\HUFe)} prioritizes tasks with the highest utility accrued on the edge $\gamma_i$, and should achieve higher net utility but possibly with fewer tasks completed.} \emph{Highest utility per execution duration First (\HPFe)} is a greedy variant of this where tasks with a higher utility on the edge per unit execution duration on the edge, \textit{i.e.}, $\frac{\beta_i - t_i \cdot \kappa_i}{t_i}$, is prioritized. This should also maximize the utility. \emph{Earliest Deadline First (\EDFe)} is deadline aware, and prioritizes tasks with earlier deadlines on the edge queue $(\delta_i)$ to maximize \textit{on-time} task completion.

\emph{Cloud-only scheduling (\CLD)} is a na\"{i}ve strategy that skips the edge and directly schedules all tasks on the cloud FaaS. This should achieve $100\%$ task completion on-time, other than for tasks with negative utility on the cloud.

\modcr{\emph{EDF (\EC)} is a hybrid strategy that uses EDF on edge and FIFO on cloud, and has been popular in literature~\cite{8917749,dai2019scheduling,7300228}. As described in Sec.5.1, tasks which cannot meet the deadline on the edge based on EDF priority are sent to the cloud for immediate execution if they have a non-negative utility. This is expected to be competitive on both on-time task completion and utility enhancement, and our \DS heuristics improve upon this. Similar to this, \emph{SJF (\EC)} is a hybrid strategy that uses Shortest Job First (SJF) on edge and FIFO on cloud, and is expected to achieve high task throughput but with reduced utility and possible deadline misses. Even tasks with a negative utility are sent to cloud in this strategy.}

\addcr{\emph{SOTA 1} is an adapted version that combines key aspects from Kalmia~\cite{fu2022kalmia} and D3~\cite{10.1145/3492321.3519576}. \textit{Kalmia}~\cite{fu2022kalmia} does QoS-aware scheduling of DNN tasks on edge servers by categorizing them into urgent and non-urgent tasks, which concept we adopt in this baseline for our VIP application. But it does task batching, which is not suitable for real-time needs. To address this, we also incorporate \textit{D3}~\cite{10.1145/3492321.3519576}, a dynamic deadline-driven DNN execution model that evaluates runtime-accuracy tradeoffs in autonomous vehicles, but without offloading to the cloud. When a new task arrives, the scheduler checks for potential deadline violations if it executes on the edge. If a violation occurs and the task is non-urgent, its deadline is increased by adding a buffer equal to $10\%$ of its current deadline value, and rescheduling is attempted. If the violation persists, the task is offloaded to the cloud. This hybrid approach ensures efficient task execution while maintaining deadline compliance in real-time scenarios.}

\addcr{\emph{SOTA 2} is another adapted version from Dedas~\cite{8941266}, tasks are inserted into the edge queue based on their \textit{expected execution time on the edge} as their priority. A global \textit{average completion time (ACT)} is maintained based on tasks that successfully complete on the edge. If inserting a task results in a schedule violation and more than one task misses its deadline, the new task is sent to the cloud. Else, we compare the updated ACT of the updated edge queue with that of the current schedule and select the one yielding the lower ACT.}

\addcr{We use the same edge and multi-threaded cloud executors as \DS and leverage the identical edge and cloud resources to ensure fairness.}

\subsection{Comparison With Baseline Algorithms}
\label{sec:exp:baselines}

\begin{figure}
  \centering
\subfloat[2D-P]{
    \includegraphics[width=0.34\linewidth]{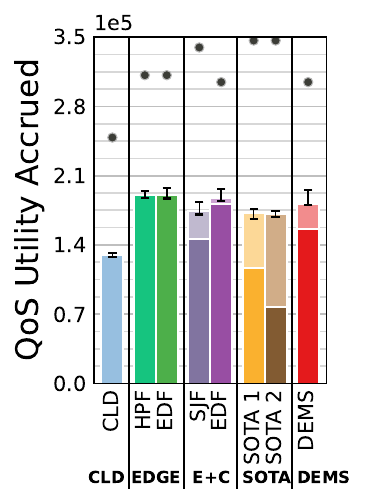}
   \label{fig:baselines:2dp}
  }\hfill
  \subfloat[2D-A]{
   \includegraphics[width=0.275\linewidth]{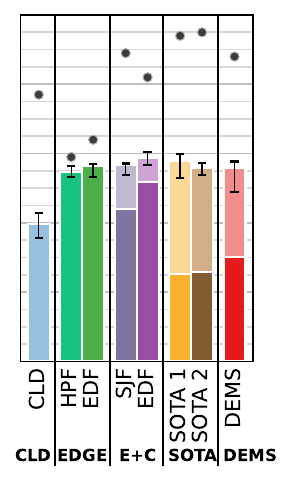}
    \label{fig:baselines:2da}
  }\hfill
  \subfloat[3D-P]{
   \includegraphics[width=0.275\linewidth]{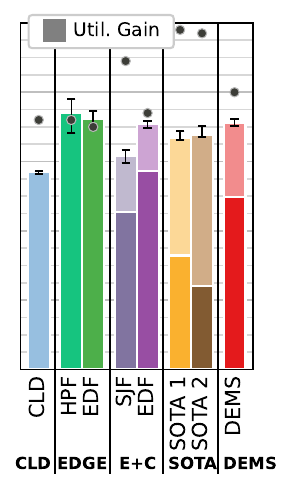}
    \label{fig:baselines:3dp}
  }\\
  \subfloat[3D-A]{
   \includegraphics[width=0.275\linewidth]{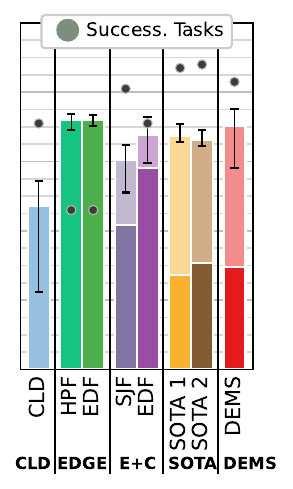}
    \label{fig:baselines:3da}
  }\hfill
  \subfloat[4D-P]{
   \includegraphics[width=0.275\linewidth]{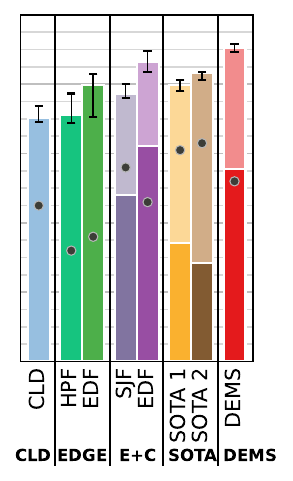}
    \label{fig:baselines:4dp}
  }\hfill
  \subfloat[4D-A]{
   \includegraphics[width=0.34\linewidth]{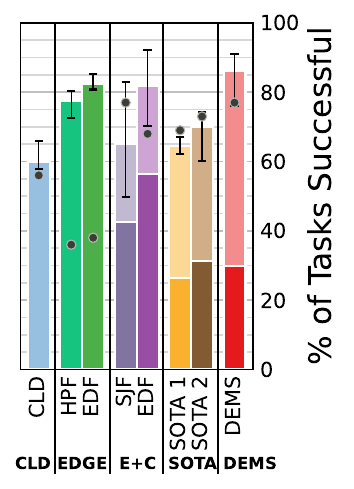}
    \label{fig:baselines:4da}
  }
    \caption{\modcr{Performance of \textit{\DS} relative to \textit{Baseline Scheduling Algorithms}. QoS utility accrued ($\times 10^5$) are shown as bars on the left Y axis. When present, the lighter-shaded stack on the top indicates utility accrued on the cloud, while the darker shade indicates utility accrued on the edge. The circle marker on the right Y axis shows the \% of tasks processed.}}
    \label{fig:baseline:bars}
\end{figure}

We compare our \DS algorithm with the \delcr{six} \modcr{seven} baselines for these application workloads that each run for $300s$ of flight time: \twoBP (2400 tasks generated per base station), \twoBA (3600 tasks), \threeBP (3600 tasks), \threeBA (5400 tasks), \modc{\fourBP (4800 tasks)} and \fourBA (7200 tasks), with the last intended to be a stress test. Each workload runs on 7 concurrent edge containers with associated drone containers (14--28) on a single host machine.

\begin{figure}
  \centering
    \includegraphics[width=0.7\columnwidth]{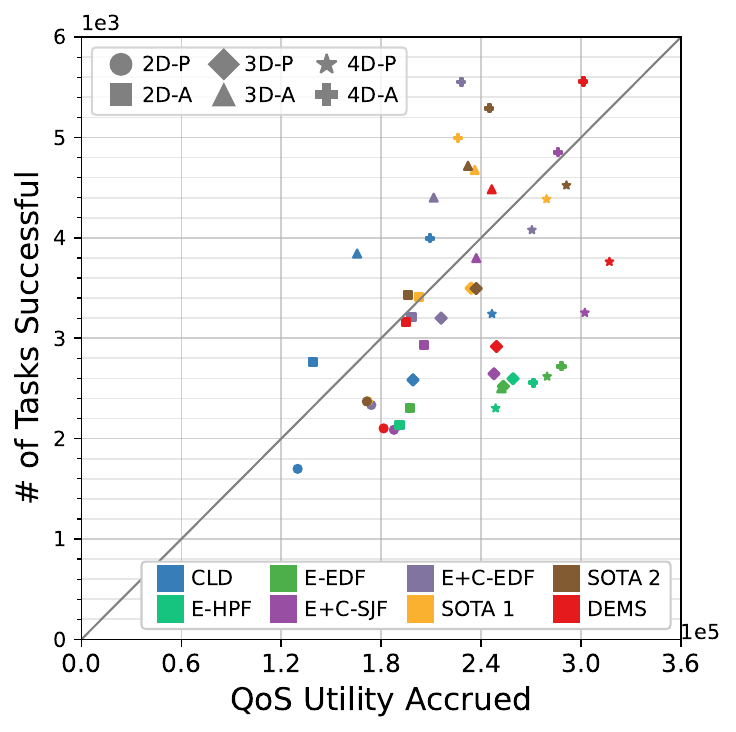}
	\caption{\modcr{Scatter plot of the \textit{Tasks Completed (Y-Axis, $\times 10^3$)} vs. \textit{QoS Utility Accrued (X-Axis, $\times 10^5$)} for \DS and the baseline scheduling algorithms, for different workloads.}}
    \label{fig:baseline:pareto}
\end{figure}

Fig.~\ref{fig:baseline:pareto} shows a scatter plot of the tasks completed (Y-axis) against the \addc{QoS} utility accrued (X-axis) by the scheduling algorithms for the different workloads. We report these for a median edge base station within the run. The diagonal indicates a balance between tasks completed and \modc{QoS} utility gained for an algorithm. \addc{Those in the top right perform the best.}

\modcr{We see \CLD's runs complete more tasks on time but suffer from a low utility. This is evident from the markers above the diagonal for active workloads. Edge-only baselines, \textit{i.e.,} \EDFe and \HPFe, on the other hand, do the opposite, completing fewer tasks but with a higher utility. \EDFe is marginally better for lighter workloads but significantly better as workload increases due to being deadline-aware. \EDFe (\EC) is competitive and closer to the diagonal, while \SJFe (\EC) is much better in terms of task completion with a lower utility. SOTA 1 and SOTA 2 are similar both in terms of tasks completion as well as utility. Our \DS strategy is able to outperform these baselines by achieving a high utility and competitive task completion rates relative to these baselines. Thanks to its heuristics, \textit{it achieves the best balance between the two competing objectives as the workload increases}.}

\modcr{We drill down into these in Fig.~\ref{fig:baseline:bars}, with the total (QoS) utility accrued (bar, left Y-axis) and the \% of tasks completed ($\bullet$, right Y-axis) shown for each algorithm and workloads for a representative edge. For the \EC variants, the utility accrued on the cloud is shown in a lighter shade for the stacked bars. The whiskers give the min and max utliity accrued across all edges and are fairly tight. So all edges perform similarly.} \modc{Among the baselines, \CLD completes the most number of tasks as the workload increases due to the ``unlimited'' capacity of the cloud. However, the utility gained is limited by the higher cost of the cloud compared to the edge (Table \ref{setup-table}). The passive workloads (\twoBP, \threeBP, \fourBP) achieve a task completion of $\approx 75\%$ since \BP gets dropped from execution on the cloud due to its negative utility. Similarly, the active workloads (\twoBA, \threeBA, \fourBA) strive to achieve a task completion of $\approx 83\%$ for 5 out of the 6 non-negative utility tasks on the cloud. Interestingly, for the most intensive \modcr{scenarios (\fourBA, \fourBP)}, only $\approx 60\%$ of task complete due to network timeouts for several tasks, from the network campus to the AWS cloud functions.}

\modcr{Since the \HPFe and \EDFe baselines use just the edge}, as we increase the workload intensity the edge saturates. \textit{E.g.}, \EDFe completes $\approx 85\%$ of its tasks for \twoBP~\modcr{(Fig.~\ref{fig:baselines:2dp})}; however this drops to only $39\%$ for \fourBA~\modcr{(Fig.~\ref{fig:baselines:4da}).} \EDFe achieves an increasing trend of utility as the workload increases. This is due to \HV, having a high utility on the edge, being prioritized by \EDFe. \addcr{As expected, \EDFe (\EC) achieves higher utility as workload increases by leveraging both edge and cloud resources to maximize on-time task processing. \SJFe (\EC) achieves a high task completion rate across all the workloads but incurs less utility as it prioritizes \DEV, which has the third lowest utility out of all DNN models. The SOTA baselines demonstrate high task completion rates, reaching up to $97\%$ for the \twoBP workload, owing to their optimized scheduling strategies. Interestingly, despite being a simple baseline, \SJFe (\EC) also attains competitive task completion rates, occasionally surpassing SOTA baselines by up to $8\%$ and achieving $77\%$ task completion rate, similar to \DS, as observed in \fourBA~(Fig~\ref{fig:baselines:4da}). However, a key observation is that while SOTA baselines and \SJFe (\EC) maintain high task completion rates, they compromise on utility by offloading all tasks—including \BP to the cloud. Although \BP completes within its deadline, it carries a negative utility on the cloud, thereby reducing overall utility. Since the utility of \BP on the cloud is relatively low (-3), the impact is minimal at lower workloads but becomes more pronounced at higher workloads. For example, in \fourBA, \DS achieves over $30\%$ higher utility compared to SOTA baselines, highlighting the trade-off between task completion and utility optimization.}
\delcr{\HUFe performs similar to others for passive workloads, but gets a performance hit on both tasks processed and utility in the active workload, since the task \DEO with an expected execution duration on edge of $739~ms$ gets the highest priority, followed by \CD with an expected duration of $563~ms$. This causes most of the tasks that are behind \DEO in the edge queue to expire, leading to a peak completion of $\approx 17\%$, \modcr{\textit{i.e.}}, just $1$ out of $6$ tasks.}

\delcr{\EC and \DS more intelligently leverage the capacity of both edge and cloud to maximize tasks processed and utility. For these plots, the number of tasks processed on the cloud is shown in a lighter shade for the stacked bars in Fig.~\ref{fig:baseline:bars}.}
\modcr{The utility accrued on edge for \DS are similar to both variants of \EC in the case of passive workloads, while they are similar to SOTA baselines for active workloads. This is because of the scoring functions that migrates more tasks to cloud based on the scores for active workloads. Additionally, we observe that the total utility of \DS is similar to edge-only variants for lower workloads, but increases for heavier workloads such as \fourBP and \fourBA, due to the heuristics kicking in. Our scheduler checks for deadline violations before inserting them into the edge queue and avoids tasks that will not be completed on time and thus waste resources. Since the \EDFe variant of \EC performs competitively with \DS in terms of both task processing and utility, we adopt \EDFe-\EC as the representative version of \EC in the next section. Hereafter, we refer to it simply as \EC while discussing the specific advantages of the \DS heuristics in the next section.}

\subsection{Benefits of Migration and Work-stealing}
\label{sec:exp:migration}

\begin{figure}
  \centering
	\includegraphics[width=0.8\columnwidth]{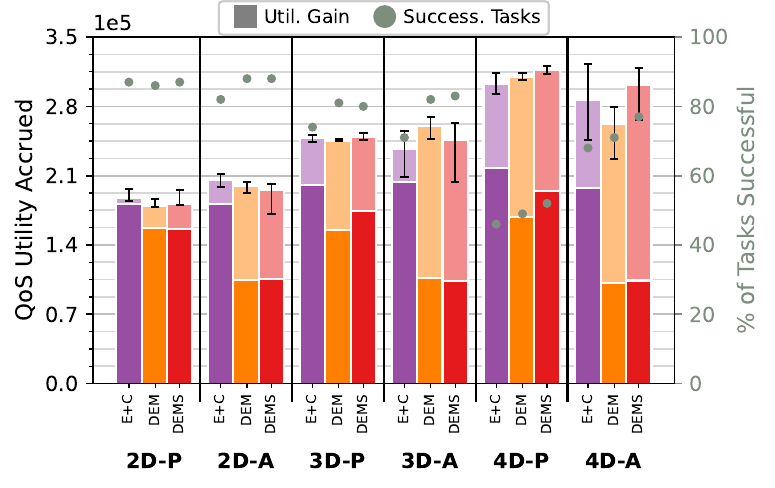}
	\caption{\modcr{Incremental benefits of \EDFcm and \DS heuristics over \EC. The lighter-shaded stack on the top indicates utility accrued on the cloud, while the darker shade indicates utility accrued on the edge.}}
    \label{fig:migration-with-stealing}
\end{figure}

We evaluate the incremental benefits of starting with \EC and including task migration (\EDFcm) and work stealing (\DS) in Fig.~\ref{fig:migration-with-stealing}. While \EC only checks the deadline of the incoming task during insertion, which may lead to the expiration of existing tasks in the edge queue, \EDFcm decides the insertion of the incoming task based on the scoring function discussed in Sec.~\ref{sec:migration}. Besides this, \DS further tries to maximize edge utilization by stealing tasks from the cloud queue when any slack accumulates in the edge queue. 

\modc{There is a significant increase in the total tasks processed for \EDFcm and \DS as the workload intensity increases. \addcr{Specifically for \fourBA, \DS processes approximately $10\%$ more tasks than \EC, and the QoS utility improves by up to $5\%$ compared to \EC and by as much as $13\%$ compared to DEM.} Since the edge resource saturates for all strategies, the number of tasks processed on the edge are comparable across them for a workload \addcr{which is reflected in the utility accrued on edge} (darker shaded bars in Fig.~\ref{fig:migration-with-stealing}). However, the cloud-processed tasks increase markedly for \EDFcm over \EC, \textit{e.g.}, by $67\%$ for \threeBA. This shows the effectiveness of the \EDFcm scoring function to insert tasks into the cloud queue. The QoS utility for \EDFcm also increases by $\approx10\%$ over \EC for \threeBA.}

\modc{
\DS incrementally improves upon \EDFcm with a rise in the total tasks processed as well as the utility due to a higher edge utilization. \modcr{\textit{E.g.}}, \DS has a peak edge utilization of $87\%$ as compared to $74\%$ for \EDFcm, for the \fourBP workload. 
The tasks that get stolen are the ones with a longer deadline while also having a shorter execution duration on the cloud and the edge. These tasks spend more time in the cloud queue before getting triggered, giving that much more time for the slack to accumulate in the edge queue to be able to pull a viable task from the cloud queue for edge execution before it expires. This also favors tasks with a negative utility on the cloud.}

\modc{
In our experiments, \BP satisfies these requirements for all the workloads. $\approx23\%$ of the successful tasks in \fourBP are stolen from the cloud queue, and are all \BP, \modcr{\textit{e.g.}}, forming $100\%$ of all successfully stolen tasks for \fourBP. In our workloads, \CD and \DEO have a higher expected execution time on the edge, which keeps the edge busy for most of the duration. Since they are a part of active workloads, we do not observe significant work stealing compared to passive workloads, as enough slack cannot be generated. Overall, we observe that DEMS achieves the highest task completion in the range of $77\% - 88\%$ across all workloads. 
These highlight the efficiency of our heuristics in responding to the diverse drone VIP workloads.}

\subsection{Adaptation to Network Variability}
\label{sec:exp:nw}

\begin{figure}[!t]
  \centering
\subfloat[Latency Variability]{
    \includegraphics[width=0.4\columnwidth]{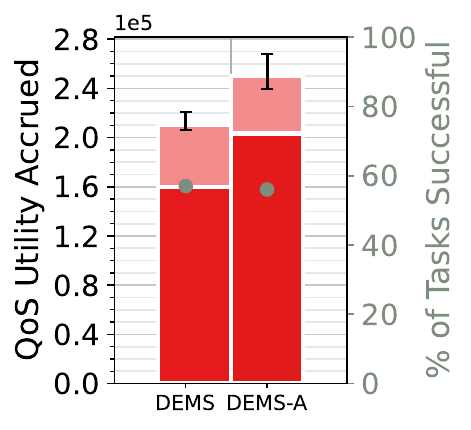}
   \label{fig:nw-adapt:lat-util:lat}
  }\qquad
  \subfloat[Bandwidth Variability]{
   \includegraphics[width=0.4\columnwidth]{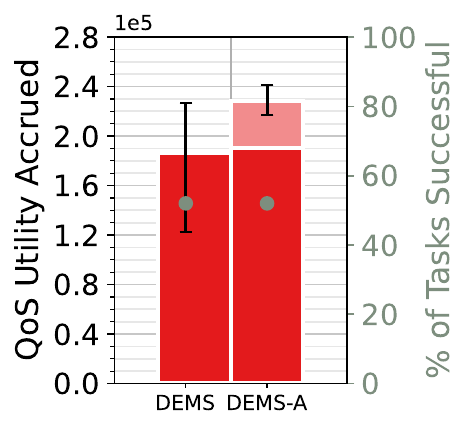}
    \label{fig:nw-adapt:lat-util:bw}
  }
  \vspace{-0.1in}
	\caption{\modcr{Benefits of \DS-A over \DS in the presence of \textit{latency (left)} and \textit{bandwidth (right) variability}.The lighter-shaded stack on the top indicates utility accrued on the cloud, while the darker shade indicates utility accrued on the edge.}}
    \label{fig:nw-adapt:lat-util}
    \vspace{-0.2in}
\end{figure}

\begin{figure}[t]
\centering

\begin{tabular}{@{}c@{}|@{}c@{}}
  \subfloat[\DS: Latency Variability]{
     \includegraphics[width=0.49\columnwidth]{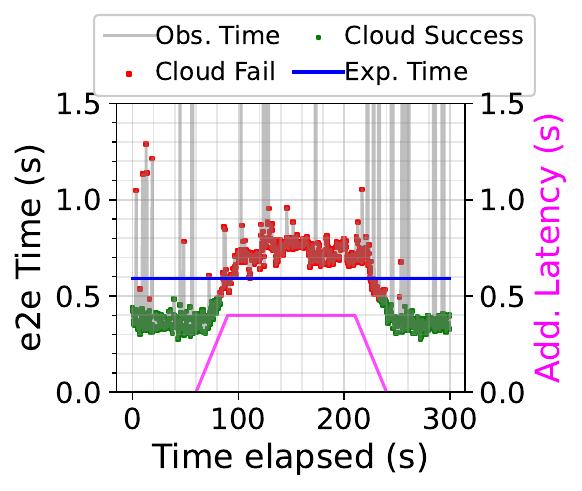}
   \label{fig:nw-adapt:lat:dems}
  }&
  \subfloat[\DS: Bandwidth Variability]{
     \includegraphics[width=0.48\columnwidth]{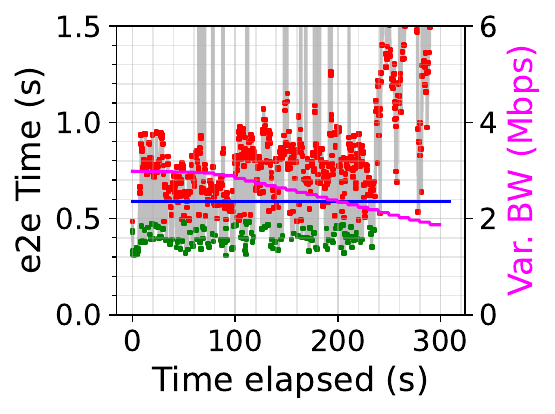}
   \label{fig:nw-adapt:bw:dems}
  }\\
  \subfloat[\DS-A: Latency Variability]{
   \includegraphics[width=0.49\columnwidth]{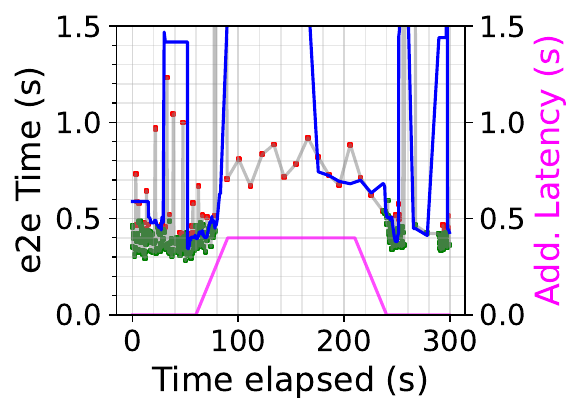}
    \label{fig:nw-adapt:lat:demsa}
  }&
  \subfloat[\DS-A: Bandwidth Variability]{
   \includegraphics[width=0.48\columnwidth]{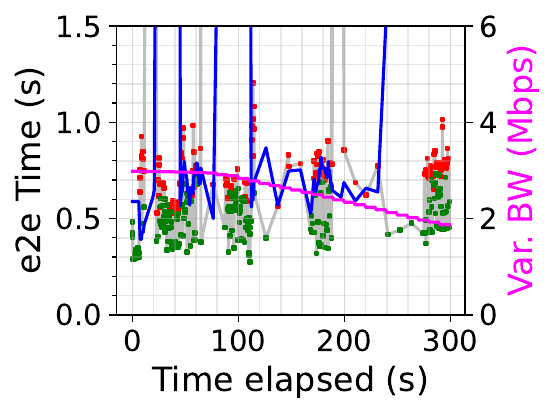}
    \label{fig:nw-adapt:bw:demsa}
  }
\end{tabular}
\vspace{-0.07in}
\caption{\modcr{End-to-end latency with \textit{\DS (top)} and \textit{\DS-A (bottom)} heuristics in the presence of \textit{variability in latency (left)} and \textit{bandwidth (right)}.}}
\label{fig:nw-adapt}
  \vspace{-0.2in}
\end{figure}

Given the mobility of drones, the network latency and bandwidth to public clouds can change over time, as illustrated in Fig.~\ref{fig:bm:nw}. Moreover, the bandwidth required by an edge increases with the number of drones connected to the base station for a VIP -- more tasks get off-loaded to the cloud with a higher workload, and more video segments need to be transferred. \modc{We evaluate the effectiveness of \textit{adapting to network variability} using \textit{\DS-A}, which extends \DS, for the \modcr{\fourBP} workload. \addcr{Figure 11 reports the task completion and total QoS utility accrued for a median edge base station within the run, with the whiskers representing the min and max QoS utility accrued across all edges.} We explicitly shape the latency and bandwidth available between the edge and the cloud to draw out the benefits.}

Specifically, we introduce a time-varying network latency $\theta$ between the edge and the cloud using Linux \texttt{tc} rules applied within the container. For easy interpretation, the latency addition follows a ``trapezium'' waveform, with $\theta$ varying between $0ms$ to $400ms$, with a linear ramp up and down at $[60s,90s)$ and $[210s,240s)$. \modc{We see from Fig.~\ref{fig:nw-adapt:lat-util:lat} that \DS-A is able to clearly improve the utility over \DS by \modcr{$19\%$} while still completing a similar number of tasks.} Drilling down into the performance of one \modcr{\DEV} task, we report the behavior of \DS (Fig.~\ref{fig:nw-adapt:lat:dems}) and \DS-A (Fig.~\ref{fig:nw-adapt:lat:demsa}) over time. The magenta line shows the $\theta$ values. The grey line is the observed cloud execution duration, with a red marker for tasks that miss their deadline and a green for those that succeed. The blue line is the expected execution time on the cloud. The \DS-A strategy is able to dynamically update its expected latency (blue line) based on the sliding window logic over the observed latency (grey line) and causes fewer cloud deadline misses; those that miss are due to the retry after the cooling period. \DS-A is also able to quickly recover once the latency drops after $210s$ and resume using the cloud once it becomes feasible.
\DS has many more cloud task failures due to the static expected execution time, which also causes negative utility. 

\delcr{We conduct a similar experiment but now vary the bandwidth between the edge container and the AWS cloud based on the simulation traces reported for the seven mobile devices within our campus(Fig.2c).}\modcr{We conduct a similar experiment but now vary the bandwidth between the edge container and the AWS cloud. We use the simulation traces from Fig.~\ref{fig:bm:nw:wan-bw}, that are generated by the seven mobile devices within our campus.}
This is much more realistic. \modc{Fig.~\ref{fig:nw-adapt:lat-util:bw} shows that here again, \DS-A outperforms \DS on utility by \modcr{$27\%$}, and is similar on on-time tasks.
We zoom into the timeline for a sample \DEV task type on a representative edge in Figs.~\ref{fig:nw-adapt:bw:dems} and ~\ref{fig:nw-adapt:bw:demsa}, with the magenta line indicating the shaped bandwidth.} The adaptation to bandwidth variability is evident with the changes in the expected deadline for \DS-A and the matching reduction in unviable tasks that it sends to the cloud.

\addcr{We have also evaluated the effectiveness of \DS-A algorithm on other workloads, that offers similar takeaways. We report them in \ref{sec:appendix-network-variability}.}

\subsection{Weak scaling of the Platform}
\label{sec:exp:scaling}

\begin{figure}[!t]
  \centering
	\includegraphics[width=0.65\columnwidth]{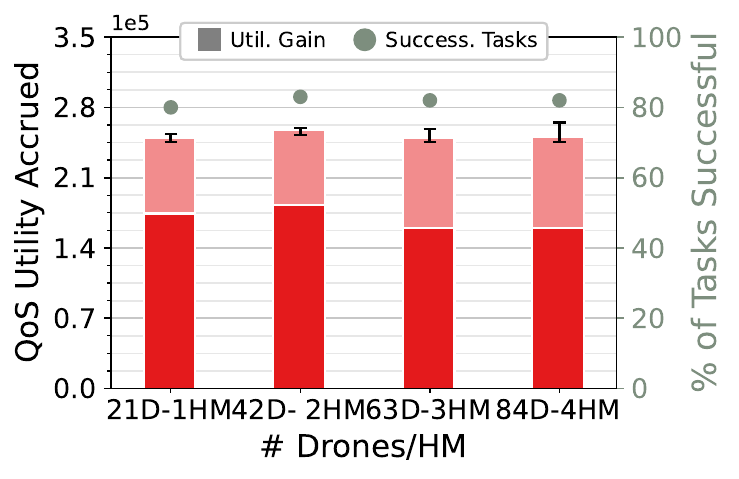}
	\caption{\modcr{Weak scaling of the platform with the number of drones (D) and host machines (HM). The lighter-shaded stack on the top indicates utility accrued on the cloud, while the darker shade indicates utility accrued on the edge.}}
    \label{fig:weakScaling}
\end{figure}

\modc{We briefly report the weak-scaling properties of our platform. The number of drones that we can support weakly scales with the number of edges, and the AWS FaaS is also able to correspondingly scale. We replicate the earlier setup of running \threeBP for $21$ drones on one host machine (HM) to $21$ drones \textit{each} on 4 host machines simultaneously. This has a peak of $84$ drones supporting 28 VIPs using 28 edge containers hosted on 4 machines.
As Fig.~\ref{fig:weakScaling} shows, the utility achieved, and the tasks completed remain almost constant with the increase in the number of drones and, correspondingly, the number of edges. \DS achieves $\approx83~\%$ task completion across all setups. We were limited by the network bandwidth from the campus to the cloud to scale more, which would not be a constraint when each edge device connects independently to a cell tower during practical deployments. \delcr{As a result, we expect the platform to be able to support a scale of 100s of VIPs for a city-scale deployment.}\addcr{As a result, we anticipate the platform to be capable of supporting a large number of users in a city-scale deployment.} }

\subsection{\addc{Guaranteeing QoE using \GS}}
\label{subsec:qoe-results}

\addc{Next, we evaluate \GS on the QoS utility and also on QoE utility that it is optimized for. \GS performs comparably to \DS for the earlier workloads, both on task completion and on QoS utility. As a result, we introduce two additional workloads, \textit{WL1} and \textit{WL2}, to highlight the benefit of \GS over \DS. The configuration for the four tasks in each workload is given in Table~\ref{gems-workload-table}.}

\addc{While the costs on the edge and cloud \modcr{($\mathcal{K},\hat{\mathcal{K}}$)} remain the same as before (Table~\ref{setup-table}), we use alternate execution durations on a sample edge ($t$) and the cloud ($\hat{t}$) to represent different compute capabilities. The deadlines for the four models also vary, to reflect alternate application needs; \MD and \CD have different configurations for the two workloads, as shown in Table~\ref{setup-table}. \addcr{We include $\Bar{\beta}$ as the QoE utility for tasks meeting QoE requirements in a window, per Eqn. 2.}.}
\addc{We configure the task completion rate ($\alpha$) for QoE to two different values, $0.9$ and $1.0$, for separate workload runs and set the window duration $\omega = 20$~seconds for all tasks. For simplicity, in these experiments, we account for the performance of these alternate edge and cloud resources by replacing the DNN execution with just a sleep function for the relevant execution duration of the tasks.}
\ysnoted{alpha=1.0 seems to be a trivial/extreme completion rate requirement}

\begin{figure}[!t]
  \centering
	\includegraphics[width=0.75\columnwidth]{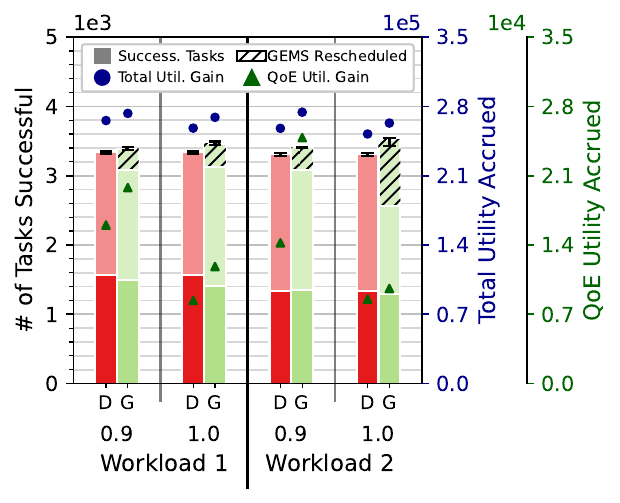}
 \vspace{-0.1in}
	\caption{\addc{ 
 Comparison of \underline{G}EMS and \underline{D}EMS for different workloads. We highlight the benefits of the heuristics using QoE utility accrued ($10^4$) shown as a triangle marker ($\blacktriangle$) on the extreme right Y axis. The lighter-shaded stack on the top indicates tasks completed on the cloud, while the darker shade indicates tasks completed on the edge. The hatched portion on the cloud processed represents GEMS rescheduled tasks. } }
    \label{fig:gatekeeper-overall}
    \vspace{-0.2in}
\end{figure}

\addc{Fig.~\ref{fig:gatekeeper-overall} shows the \textit{tasks completed} (bar, left Y-axis), \textit{QoE utility} accrued (green $\blacktriangle$, far right Y-axis) and the \textit{total utility} accrued (blue $\bullet$, right Y-axis) for \DS (left) and \GS (right) on the two workloads and using both $\alpha$ threshold completion rates, on a representative edge.
Additionally, we show the number of tasks successfully processed due to \GS's greedy re\-schedul\-ing as a hatched stack above the cloud-processed stack.}

\begin{table}[t]
\centering
\caption{\addc{Workload config. for DNN tasks on a sample edge to evaluate \GS.}}
\label{gems-workload-table}
\vspace{-0.07in}
\footnotesize
\begin{tabular}{l||rrrr||@{}c@{}|@{}c@{}}
\toprule
\textbf{DNN} &~~~ $\bar{\beta}$ &~~~ $\delta$ &~~~ $t$ &~~~ $\hat{t}$ & ~WL1~ & ~WL2~\\
\midrule
\HV & 360 & 400 & 100 & 200 & \checkmark & \checkmark\\ 
\hline
\DEV & 420 &  600 & 300 & 400 & \checkmark & \checkmark \\
\hline
$\MD$-\textit{WL1} & 480 & 1000 & 200 & 300 & \checkmark & $\times$ \\  
\hline
$\CD$-\textit{WL1} & 600 & 800 & 650 & 750  & \checkmark & $\times$\\
\hline
$\MD$-\textit{WL2} & 480 & 800 & 200 & 300 & $\times$ & \checkmark \\  
\hline
$\CD$-\textit{WL2} & 600 & 1000 & 750 & 950  & $\times$ & \checkmark\\
\bottomrule 
\end{tabular}
\end{table}

\begin{figure}[!t]
  \centering
  \subfloat[Tasks processed per DNN Model]{
    \includegraphics[width=0.85\columnwidth]{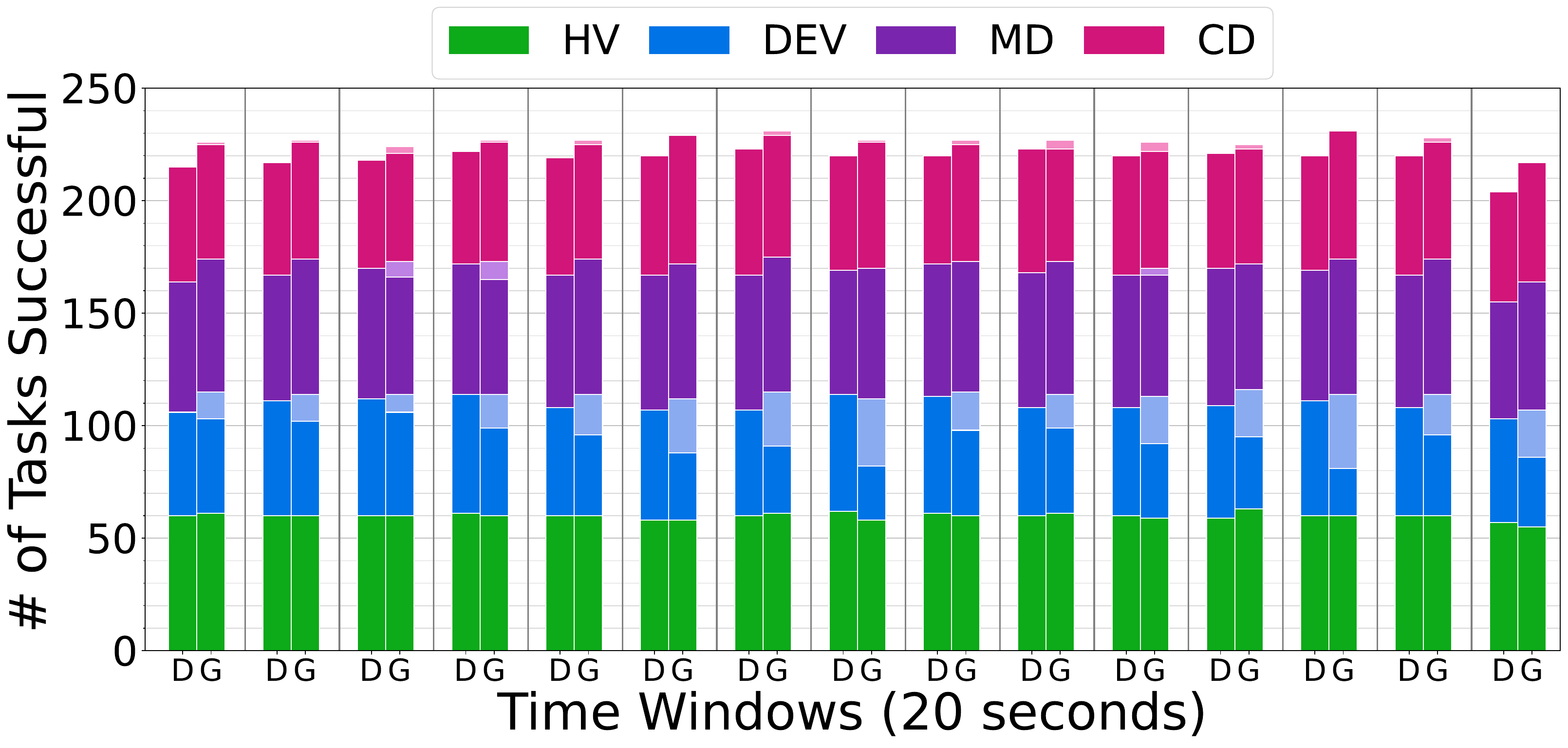}
   \label{fig:gatekeeper-taskprocessed}
  }\\
    \vspace{-0.07in}
  \subfloat[Total utility accrued per DNN Model]{
    \includegraphics[width=0.85\columnwidth]{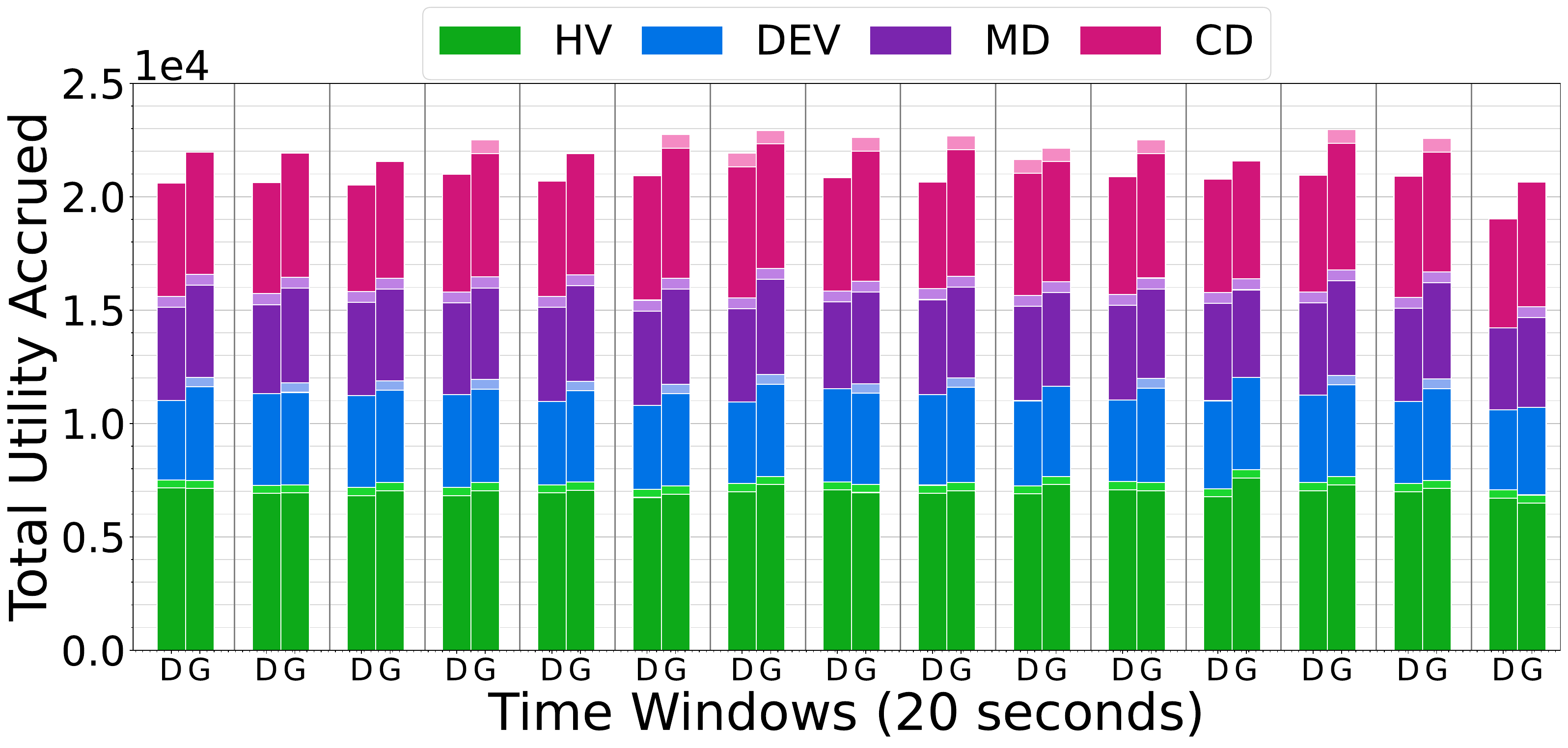}
   \label{fig:gatekeeper-utility}
  }
    \vspace{-0.07in}
  \caption{\addc{\textit{Tasks processed} and \textit{Total utility accrued} per DNN Model for \underline{G}EMS and \underline{D}EMS (Y axis) for a series of tumbling window of $\omega=20$~seconds each (X axis). The lighter-shaded stack indicates tasks completed due to \GS rescheduling in (a) and QoE utility accrued in (b).}}
    \vspace{-0.2in}
    \end{figure}

\addc{
\GS outperforms \DS for both the workloads and both the completion rates, in terms of task completion and also on total utility accrued, improving these metrics by up to $7\%$. 
For WL1, \GS shows a $24\%$ and $40\%$ increase in QoE utility over \DS for $\alpha=0.9$ and $1.0$, respectively. This is because \DEV and \CD tasks expire on the edge in \DS, but their impending expiration is detected ahead of time by \GS and rescheduled to the cloud.
These constitute a total of $306$ and $342$ rescheduled tasks by \GS for $\alpha=0.9$ and $1.0$, a gain of $1.5\%$ and $3.8\%$ of tasks.
\modcr{For WL2, we see an increase of $75\%$ and $13\%$ in the QoE utility for \GS for $\alpha=0.9$ and $1.0$, respectively. Also, for $\alpha=1.0$, we have $961$ out of $2230$ cloud-processed tasks due to rescheduling by \GS.} 

These benefits again occur due to \GS preemptively identifying the low task completion rate for \DEV and \CD in a window and pushing them to the cloud from the edge queue, where possible. Models with a lower execution duration and higher deadline are best suited for this as they offer \GS an opportunity to proactively act. \DEV and \MD in WL1 fall in this category, while only \DEV falls in this category for WL2. Due to this, \DEV constitutes $\approx 86\%$ of the \GS rescheduled tasks for WL2 while for WL1, we observe $87\%$ \DEV tasks in \GS rescheduled for $\alpha=0.9$ while only $56\%$ for $\alpha=1.0$, and $36\%$ are constituted by \MD.}   

\addc{We further zoom into WL1 with $\alpha=0.9$ to show the split of tasks processed (Fig.~\ref{fig:gatekeeper-taskprocessed}) and utility accrued (Fig.~\ref{fig:gatekeeper-utility}) for a series of $15$ time windows of $\omega=20s$ duration over a $300s$ experiment run, on a representative edge. \HV, \CD and \MD achieve similar task completion rate for both \DS and \GS. But \DEV using \DS finishes only $\approx 50$ out of $60$ tasks in a window whereas with \GS, it executes $\approx 55$ tasks out of $60$ in a window. This enhances the task completion rate from $\approx 80\%$ for \DS to $\geq 90\%$ for \GS and increases the QoE utility by $5\%$--$10\%$ for $12$ of the $15$ windows. These empirically validate the benefits of \GS in achieving better QoE and total utility than \DS.}

\subsection{\addc{Practical Validation in Realistic Experiments} \label{subsec:hardware-validation}}

\begin{figure}[!t]
  \centering
\subfloat[\addc{Proxy} VIP as seen by Tello Drone]{
    \includegraphics[width=0.24\columnwidth]{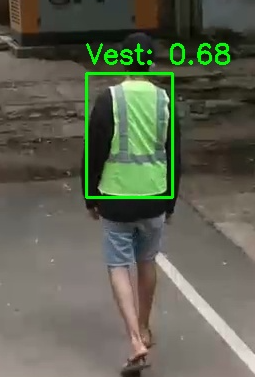}
   \label{fig:tello-view}
  } \qquad
  \subfloat[\addc{Proxy} VIP \addc{autonomously} followed by the Tello \addc{using inferencing}]{
   \includegraphics[width=0.35\columnwidth]{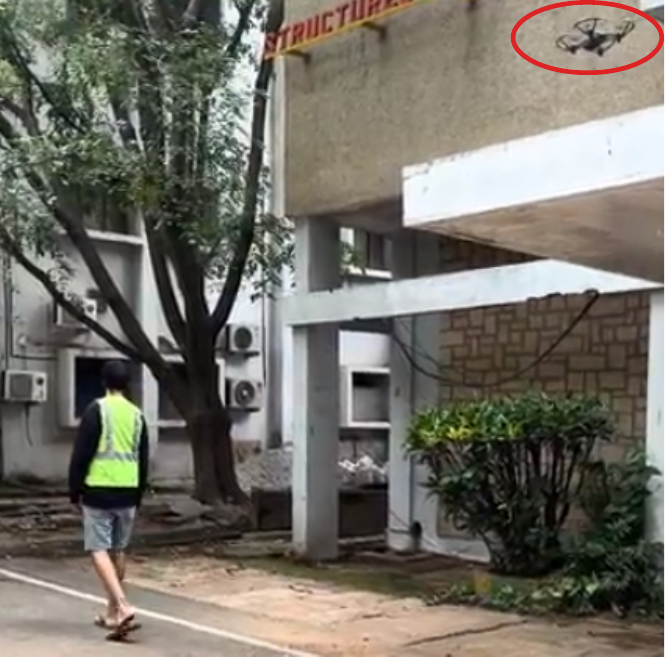}
    \label{fig:tello-pic}
  }
  \vspace{-0.07in}
	\caption{\modc{Field-validation using our inferencing architecture for VIP domain. Full video is shown supplementary material in~\ref{sec:supplementary-video}.}}
    \label{fig:validation}
    \vspace{-0.2in}
\end{figure}

\addc{Lastly, we demonstrate the practical benefits of our scheduling strategies and inferencing architecture for the VIP domain described in \S~\ref{sec:app-based-framework} through a field-validation in our university campus.
We use a \textit{Tello nano quad-copter} (Fig.~\ref{fig:tello} in Appendix), which has an onboard $720$p HD monocular camera that generates live video feeds at $30$ FPS and weighs just $80$~g with battery, along with the latest generation \textit{Nvidia Jetson Orin Nano} edge device~\cite{orin-tech-specs} (Fig.~\ref{fig:orin-nano}). 
The Orin Nano interfaces with the Tello drone over WiFi using a $150$~Mbps TP-link 802.11n USB adapter with a range of $100m$ (clear line of sight) to send live video feeds from the drone camera to the edge, and to send control commands from the edge to the drone. }

\begin{figure}[t]
\centering
  \subfloat[\DS relative to baselines \\for hardware experiments.]{
    \includegraphics[width=0.56\columnwidth]{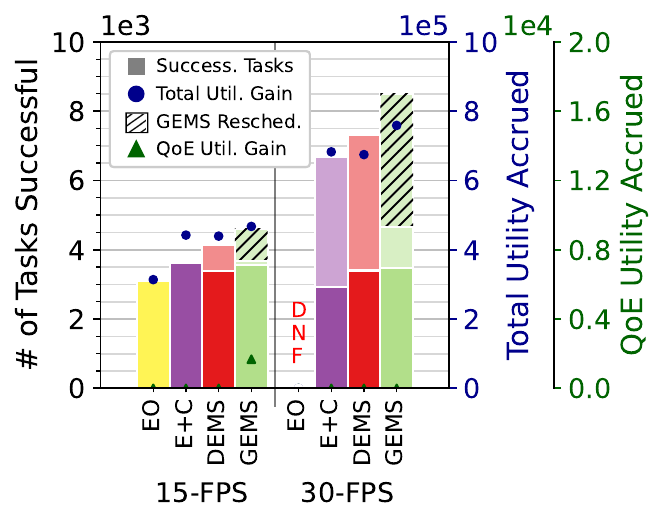}
   \label{fig:hardware-baselines}
  }\hfill
  \subfloat[Post Processing Latencies \\for various DNN Models]{
    \includegraphics[width=0.4\columnwidth]{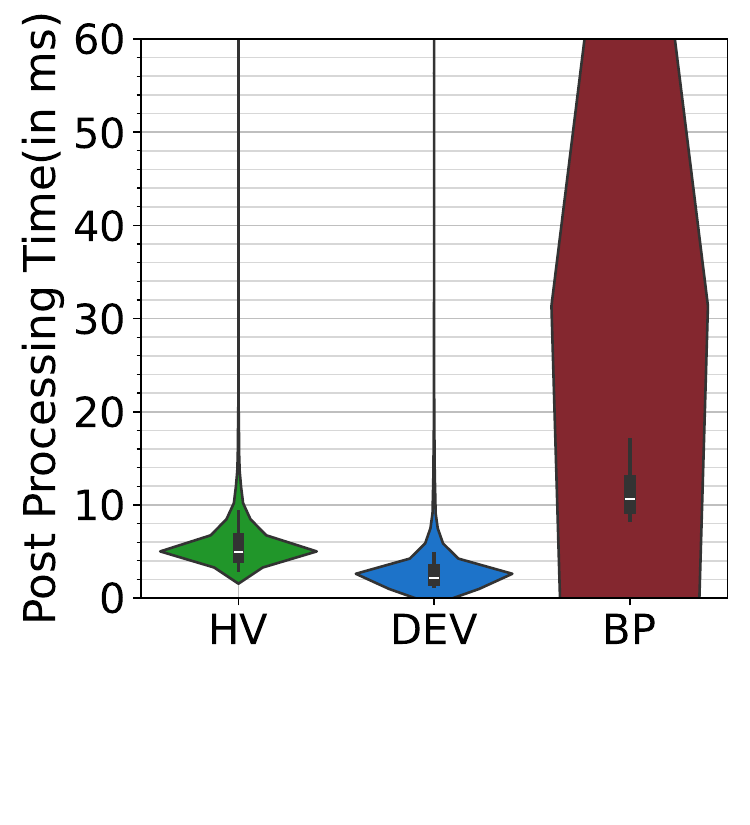}
    \label{fig:hardware-postprocessing}
  }
  \vspace{-0.07in}
\caption{\addc{Performance of \GS scheduling strategies and integration of post-processing logic on hardware. In the left figure, the lighter-shaded stack on the top indicates tasks completed on the cloud, while the darker shade indicates tasks completed on the edge.}}
\label{fig:hardware-performance}
\vspace{-0.1in}
\end{figure}

\addc{The scheduler runs on the Orin Nano, along with the Ocularone application logic (Fig.~\ref{fig:app-based-framework}).
The drone is setup to follow a proxy VIP wearing the hazard vest and walking around our campus. Three DNN models are active, \HV, \DEV and \BP, to consume the drone's video feed and we schedule the inferencing using the \DS and \GS heuristic and other baselines for comparison. The scheduler creates one \HV task for every incoming frame (\modcr{\textit{e.g.}}, 30 FPS), and a task for every $3^{rd}$ frame of \DEV and \BP models (\modcr{\textit{e.g.}}, 10 FPS). This is consistent with our prior validation without using the scheduler~\cite{suman2023chi}. The $99^{th}$ percentile of expected execution times per frame for each model on the \textit{Orin Nano} and their normalized costs, \modcr{($t, \mathcal{K}$)}, is: (\HV: $49ms,~1$; \DEV: $50ms,~1$; \BP: $72ms,~1$). The rest of the configuration parameters are retained from Table~\ref{setup-table} for Jetson Nano. The scheduler executes these tasks as per the scheduling heuristic and returns the output of model inferencing. The output of \HV (Fig.~\ref{fig:tello-view}) is used by the $PD$ controller running on the Orin Nano to generate commands to the drone to autonomously follow the proxy VIP while maintaining a constant distance of $3m$ (Fig.~\ref{fig:tello-pic}). The latency to send the control commands from the Orin Nano to the drone is a negligible $\approx 2~ms$. We also report the post-processing latencies of the DNN models in Fig.~\ref{fig:hardware-postprocessing}, where the median latency across all experiments is $4ms, 2ms$ and $10ms$ for \HV, \DEV and \BP, respectively, indicating that the overhead of post-processing the inferencing outputs is minimal. This allows for real-time tracking of the VIP and their pose estimation for the situation awareness. The full demonstration video is given in~\ref{sec:supplementary-video}.}

\begin{figure}[t]
  \centering
\subfloat[\modcrr{Jerk for drone's motion at 15 FPS}]{
    \includegraphics[width=0.48\columnwidth]{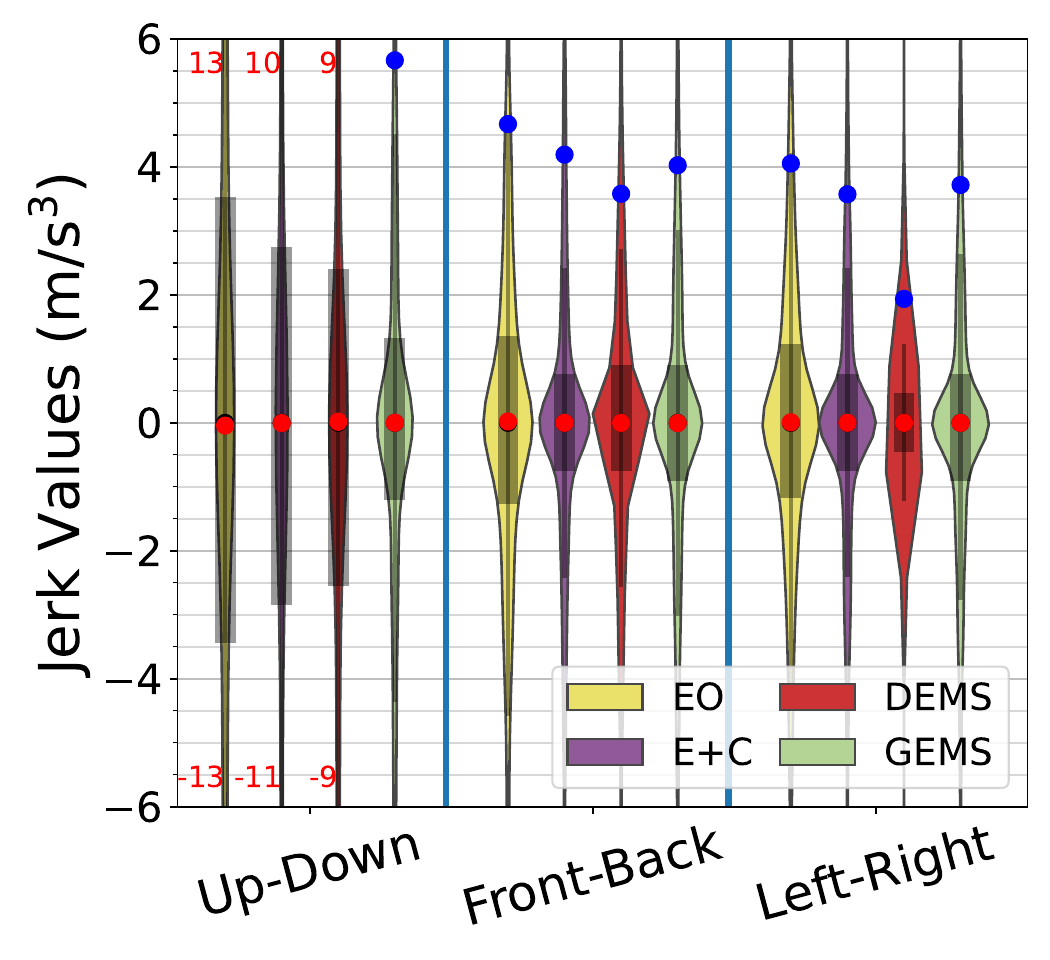}
   \label{fig:jerk-error-revised}
  }\hfill
  \subfloat[\modcr{Yaw error for drone's motion}]{
   \includegraphics[width=0.48\columnwidth]{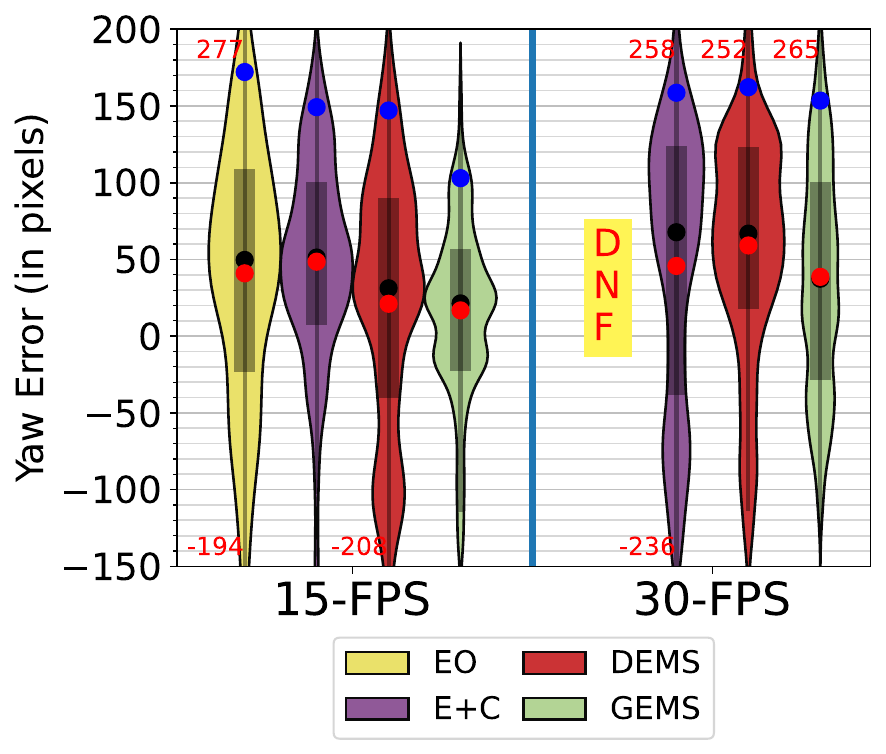}
    \label{fig:yaw-error-revised}
  }
  \vspace{-0.1in}
  \caption{\modcrr{Drone mobility error metrics from field experiments. Mean values are represented by a red dot ($\color{red}\bullet$), median values by a black dot ($\color{black}\bullet$) and the $95^{th}$ percentile values by a blue dot ($\color{blue}\bullet$). }}
    \label{fig:practical validation-revised}
    \vspace{-0.2in}
\end{figure}

\addc{We report results for video feeds sent at $15$ and $30$ FPS, and compare \GS with a task completion rate of $\alpha=1.0$ for a window duration of $\omega=20s$, against \DS and two baselines: (1) Edge Only (EO) scheduling using only the edge accelerator, and (2) Edge + Cloud (\EC), using both the edge and AWS Lambda cloud FaaS. Each experiment runs for around $3.5$~minutes. As expected, Fig.~\ref{fig:hardware-baselines} confirms that \GS achieves the best performance on overall task completion and total utility. It completes $48\%$ more tasks compared to EO at $15$~FPS, and $27\%$ more tasks compared to \EC at $30$~FPS. We see the utility improve by $11\%$ due to more tasks being rescheduled by \GS to the cloud at $30$~FPS even though this setup does not accrue the full QoE utility due to the strict $1.0$ completion rate. Interestingly, the experiment with EO at 30 FPS fails as the \HV tasks expire due to queuing delays and the drone is unable to fly beyond a few seconds since it does not receive any PID control commands; as a safety measure, it lands. This is shown as \textit{Did Not Finish (DNF)}.}

\addc{Finally, we examine how the QoE metrics we define translate to qualitative metrics for the drone mobility domain. Fig.~\ref{fig:jerk-error-revised} reports a distribution of the \textit{jerk} ($J$), which is the derivative of the drone's acceleration with respect to time in the $x$ (front-back), $y$ (left-right), and $z$ (up-down) directions, given by $J(t) = \frac{d\,a(t)}{dt}$. A lower value of jerk indicates that the drone's trajectory is smoother, with $[-1.0,~2.0]m/s^3$ being an acceptable range for a mini-quadrotor~\cite{shupeng2018safe}, and $\pm0.9m/s^3$ for passenger comfort in autonomous vehicles~\cite{bae2020self}.}

\modcrr{In Fig.18(a), \GS consistently exhibits a minimal \textit{jerk} in all three directions, staying within $\approx \pm 1.0$. This is particularly seen along the $z$-axis (Up-Down), where the $95^{th}$ percentile jerk for \GS (blue dot) is the lowest at $5.67~\text{m/s}^3$, compared to values exceeding $6~\text{m/s}^3$ in other strategies. 
In Fig. 18(b), the differences in yaw error are more pronounced. \GS consistently achieves lower mean (red dot), median (black dot), and $95^{th}$ percentile errors (blue dot) at both $15$ and $30$ FPS. At $15$ FPS, \GS demonstrates up to a $41\%$ reduction in the $95^{th}$ percentile yaw error compared to EO. Similarly, for $30$ FPS, the median error is reduced by up to $42\%$ with \GS in comparison to \EC. This is a result of the PD controller's adaptation to the drone's camera to ensure that VIP remains at the center of the frame. These lower errors demonstrate a better adaptation using \GS's heuristic compared to others.}

\modcrr{We see that the Front-Back and Left-Right do not show much jerk for all algorithms, compared to the better performance of \GS for yaw motion and height changes (Up-Down). This is a consequence of the workload we evaluate this for, where the drone is following the VIP through some sharp turns and stairs, causing the PD control to change the yaw and height often. Hence, the scheduler's effect on jerk is seen more along these motion axes.}

\vspace{-0.3cm}
\section{\addcr{Discussion and Limitations}} \label{sec:discussion}
\vspace{-0.2cm}

\modcrr{Using drones to assist visually impaired individuals in urban environments is a promising direction but comes with several legal, regulatory, and technical challenges. Most countries enforce strict safety and privacy regulations, restricting drone operations near sensitive areas and requiring adherence to national aviation standards. E.g., the USA follows FAA's Part 91 regulations, which outline general operating and flight rules, including restrictions on night operations; the EU applies a unified drone regulation framework under EASA with some local deviations; India mandates No Permission No Takeoff (NPNT) protocols with certain relaxations for drones under 250 grams; and China imposes stringent restrictions in urban areas.}

\modcrr{Privacy concerns are particularly significant due to the use of onboard cameras, which may inadvertently record individuals or private properties. This requires obtaining consent and complying with local data protection laws, such as the General Data Protection Regulation~\cite{gdpr_overview}. Moreover, bystanders may feel uncomfortable around drones in public spaces and may interfere with their operations. While hazard vests are helpful for identification, they face limitations such as glare or misidentification in crowded scenes. Alternative approaches, such as using QR codes, alphanumeric tags, or wearable accessories like caps and Bluetooth tethers, could improve reliability and better support practical adoption.}

\modcrr{There are also several system-level challenges affecting drone-assisted navigation. Edge hardware has limited compute capacity, restricting the number of concurrent DNN models and drone clients it can support. Excessive cloud-bound inference requests can lead to queuing delays, which are partially mitigated by FaaS techniques, though these are constrained by the memory footprint allowed for such functions. Additionally, energy consumption is not currently modeled as an optimization goal or constraint. This can limit drones and edge devices running off battery with restriced capacities. Stable cellular or Wi-Fi connectivity is essential for effective offloading to the cloud when flying. Lightweight drones such as the Tello, for instance, have limited flight endurance ($\approx13$ minutes) and are vulnerable to environmental factors like wind.}

\modcrr{Further, the performance of task scheduling algorithms depends on the characteristics of the workload. Our system model assumes the use of a captive private edge accelerator co-located with the drone. This setup is typically more cost-effective when actively used, compared to public cloud services. However, if public edge services such as AWS Lambda@Edge or Akamai EdgeWorkers are employed, the cost of edge inference is likely to exceed that of cloud inference, and the scheduling optimizations we propose may not yield the same benefits.}

\modcrr{Addressing these constraints will be essential for enhancing the practical viability of drone-assisted navigation systems for visually impaired persons.}

\vspace{-0.3cm}
\section{Conclusion and Future Work} \label{sec:conclusion}
\vspace{-0.2cm}
\modc{In this article, we have proposed an inferencing architecture to support a novel application platform for assisting VIPs using buddy drones connected to edge accelerators. We have defined a deadline-driven scheduling problem for streams of DNN inferencing tasks to execute on edge and cloud resources, to help maximize QoS utility, domain-driven QoE utility and tasks completion within the deadline. Our \DS heuristic achieves superior QoS benefits on utility and task counts relative to baselines, based on detailed evaluations using realistic workloads, using container-based edge and AWS public cloud deployments. \DS-A adapts to cloud variability and greatly reduces the deadline misses on the public cloud over WAN to improves QoS utility.}
\addc{Further, \GS builds upon these to target the QoE goals and improves the QoE utility for different DNN models, which also translates to real-world benefits such as a smoother drone navigation.} \addcr{The \GS scheduling algorithm uses latency, cost and task completion rate as QoE goals for the analytics to make decisions. Later, we will explore the automated definition of these quality metrics based on SLAs defined by the VIP users or by drone service providers to meet specific safety and response requirements for their applications.}

Our platform offers apps access to an observe, orient, decide and act (OODA) loop to leverage multiple personalized drones to support human activity in an easily programmable manner.
This can be leveraged beyond supporting VIPs to other latency-sensitive drone-based applications as well, such as intelligent traffic monitoring, supporting racers in a marathon, assisting safety officers, etc. \addc{We demonstrate the integration of our scheduling platform with the Ocularone application on real hardware and show that \DS offers the smoothest navigation compare to the baselines.}

In the future, we propose to generalize the application \modc{platform to other drone mobility applications and perform complex tasks cooperatively.}
\delc{We also plan to validate our edge and cloud runtime for large drone fleets under real-world conditions.}%
The scheduling problem will also be redefined if the size of the video segments or their frame rates should vary based on the accuracy needs of the DNN models for different applications.
\delc{, \modcr{\textit{e.g.}}, requiring higher FPS when the VIP is running or there is fast-moving traffic.}%
\delc{Further, consumer drones and edge accelerator devices are constrained by their battery life. \modcr{\textit{E.g.}}, the Ryze Tello has a flight time of $13$~minutes while the larger Mavic Mini can fly for $30$~minutes. Edge accelerators like the Nvidia Jetson Orin Nano have a peak power of $15~W$ and can be powered using a compact battery power bank.}\modcrr{Updating the expected execution duration for tasks already in the cloud queue is an interesting problem and offers valuable direction for future investigation.}

\modcrr{Drones operate on rechargeable batteries, with energy consumption varying by drone model and outdoor conditions.} As drones start having longer endurance, it will open up interesting problems with energy-aware scheduling on the edge \addc{to jointly optimize for navigation and analytics}. \addcr{Further, the problem statement can be extended to explore energy consumption as a primary optimization goal or constraint. Machine learning algorithms can further enhance scheduling by predicting model selection 
to balance energy usage and DNN accuracy.}
\delc{Here, the edge has not just a limited instantaneous compute capacity but a bounded cumulative capacity due to the battery limits, and this has to be factored in when deciding the placement on edge or cloud. A graceful degradation in application capability is also required when the battery runs low.}

\vspace{-0.5cm}
\section*{Acknowledgments}
\vspace{-0.2cm}
The authors would like to thank members of DREAM:Lab, including \modc{Bhavani A Madhabhavi, Sahil Sudhakar and Rajdeep Singh} for their assistance with this paper. 
The first author was supported by a Prime Minister's Research Fellowship (PMRF) from the Government of India.

\vspace{-0.5cm}
\bibliographystyle{elsarticle-num} 
\bibliography{ccgrid}

\begin{thebibliography}{10}
\expandafter\ifx\csname url\endcsname\relax
  \def\url#1{\texttt{#1}}\fi
\expandafter\ifx\csname urlprefix\endcsname\relax\def\urlprefix{URL }\fi
\expandafter\ifx\csname href\endcsname\relax
  \def\href#1#2{#2} \def\path#1{#1}\fi

\bibitem{whoStats}
{World Health Organization}, \href{https://www.who.int/news-room/fact-sheets/detail/blindness-and-visual-impairment}{Blindness and visual impairment fact sheets} (August 2023).

\bibitem{suman2023chi}
S.~Raj, S.~Padhi, Y.~Simmhan, Ocularone: Exploring drones-based assistive technologies for the visually impaired, in: Extended Abstracts of the CHI Conference on Human Factors in Computing Systems, 2023.

\bibitem{faa}
C.~of~Federal~Regulations, \href{https://www.ecfr.gov/current/title-14/chapter-I/subchapter-F/part-91}{General Operating and Flight Rules (14 CFR Part 91)}, Tech. rep., Federal Aviation Administration (2024).

\bibitem{easa}
E.~U. A.~S. Agency, \href{https://www.easa.europa.eu/en/light/topics/drones}{Easy Access Rules for Unmanned Aircraft Systems (Regulations (EU) 2019/947 and 2019/945)}, Tech. rep. (2022).

\bibitem{dgca}
M.~of~Civil~Aviation, \href{https://www.dgca.gov.in/digigov-portal/Upload?flag=iframeAttachView&attachId=151915836}{Accessibility Standards and Guidelines for Civil Aviation}, Tech. rep., India (2023).

\bibitem{9128519}
C.~Christodoulou, P.~Kolios, Optimized tour planning for drone-based urban traffic monitoring, in: IEEE 91st VTC2020-Spring, 2020.

\bibitem{benarbia2021literature}
T.~Benarbia, K.~Kyamakya, A literature review of drone-based package delivery logistics systems and their implementation feasibility, Sustainability 14~(1) (2021) 360.

\bibitem{KUMAR20211}
A.~Kumar, et~al., A drone-based networked system and methods for combating coronavirus disease (covid-19) pandemic, Future Generation Computer Systems 115 (2021).

\bibitem{hiebert2020application}
B.~Hiebert, E.~Nouvet, V.~Jeyabalan, L.~Donelle, The application of drones in healthcare and health-related services in north america: A scoping review, Drones 4~(3) (2020) 30.

\bibitem{avila2015dronenavigator}
M.~Avila, M.~Funk, N.~Henze, Dronenavigator: Using drones for navigating visually impaired persons, in: ACM SIGACCESS, ASSETS, 2015.

\bibitem{avila2017dronenavigator}
M.~Avila~Soto, M.~Funk, M.~Hoppe, R.~Boldt, K.~Wolf, N.~Henze, Dronenavigator: Using leashed and free-floating quadcopters to navigate visually impaired travelers, in: ACM SIGACCESS, ASSETS, 2017.

\bibitem{orin-tech-specs}
NVIDIA, \href{https://developer.nvidia.com/downloads/assets/embedded/secure/jetson/orin_nano/docs/jetson_orin_nano_devkit_carrier_board_specification_sp.pdf}{Jetson Orin Nano Developer Kit Carrier Board Specification}, Tech. rep., NVIDIA Corporation (2024).

\bibitem{kuriakose2022tools}
B.~Kuriakose, R.~Shrestha, F.~E. Sandnes, Tools and technologies for blind and visually impaired navigation support: a review, IETE Technical Review 39~(1) (2022) 3--18.

\bibitem{al2016exploring}
M.~Al~Zayer, S.~Tregillus, J.~Bhandari, D.~Feil-Seifer, E.~Folmer, Exploring the use of a drone to guide blind runners, in: ACM SIGACCESS, ASSETS, 2016.

\bibitem{drone-first-aid}
A.~Momont, \href{https://www.tudelft.nl/en/ide/research/research-labs/applied-labs/ambulance-drone/}{Ambulance Drone at TU Delft}, Tech. rep., TUDelft (2014).

\bibitem{ZENG20201028}
J.~Zeng, L.~T. Yang, M.~Lin, H.~Ning, J.~Ma, A survey: Cyber-physical-social systems and their system-level design methodology, FGCS 105 (2020).

\bibitem{10.1145/3570604}
{S.K. Prashanthi}, S.~A. Kesanapalli, Y.~Simmhan, Characterizing the performance of accelerated jetson edge devices for training deep learning models, Proc. ACM Meas. Anal. Comput. Syst. (2022).

\bibitem{7300228}
A.~Yamaguchi, Y.~Nakamoto, K.~Sato, Y.~Watanabe, H.~Takada, Edf-pstream: Earliest deadline first scheduling of preemptable data streams -- issues related to automotive applications, in: IEEE Intl. Conf. on Embedded and Real-Time Computing Syst. \& Applications, 2015.

\bibitem{8685783}
J.~Wang, Z.~Feng, Z.~Chen, S.~A. George, M.~Bala, P.~Pillai, S.-W. Yang, M.~Satyanarayanan, Edge-based live video analytics for drones, IEEE Internet Computing (2019).

\bibitem{guo2019uav}
H.~Guo, J.~Liu, Uav-enhanced intelligent offloading for internet of things at the edge, IEEE Transactions on Industrial Informatics (2020).

\bibitem{khochare:ton:2024}
A.~Khochare, F.~B. Sorbelli, Y.~Simmhan, S.~K. Das, Improved algorithms for co-scheduling of edge analytics and routes for uav fleet missions, IEEE/ACM Transactions on Networking 32~(1) (2024).

\bibitem{10171496}
S.~Raj, H.~Gupta, Y.~Simmhan, Scheduling dnn inferencing on edge and cloud for personalized uav fleets, in: IEEE/ACM 23rd CCGrid, 2023.

\bibitem{MAHMUD2019190}
R.~Mahmud, S.~N. Srirama, K.~Ramamohanarao, R.~Buyya, Quality of experience (qoe)-aware placement of applications in fog computing environments, Journal of Parallel and Distributed Computing 132 (2019).

\bibitem{9723632}
Y.~Mao, et~al., Differentiate quality of experience scheduling for deep learning inferences with docker containers in the cloud, IEEE Transactions on Cloud Computing 11~(2) (2023) 1667--1677.

\bibitem{islam2019developing}
M.~M. Islam, M.~Sheikh~Sadi, K.~Z. Zamli, M.~M. Ahmed, Developing walking assistants for visually impaired people: A review, IEEE Sensors Journal (2019).

\bibitem{weWalk}
{WeWALK Limited UK}, \href{https://wewalk.io/en/}{WeWalk: Enhancing the mobility of visually impaired people} (August 2020).

\bibitem{aiPoweredBackpack}
D.~Brown, \href{https://www.washingtonpost.com/technology/2021/03/25/innovations-ai-backpack-blind/}{Researchers design an AI-powered backpack for the visually impaired} (2021).

\bibitem{bai2017smart}
J.~Bai, S.~Lian, Z.~Liu, K.~Wang, D.~Liu, Smart guiding glasses for visually impaired people in indoor environment, IEEE Transactions on Consumer Electronics (2017).

\bibitem{lookout}
A.~Al-Heeti, Google expands lookout app for people who are blind or vision-impaired, Tech. rep., CNet (Aug. 2020).

\bibitem{nasralla2019computer}
M.~M. Nasralla, I.~U. Rehman, D.~Sobnath, S.~Paiva, Computer vision and deep learning-enabled uavs: Proposed use cases for visually impaired people in a smart city, in: Computer Analysis of Images and Patterns, Springer International Publishing, 2019.

\bibitem{8622052}
H.~Tianfield, Towards edge-cloud computing, in: IEEE International Conference on Big Data (Big Data), 2018.

\bibitem{varshney2020characterizing}
P.~Varshney, Y.~Simmhan, Characterizing application scheduling on edge, fog, and cloud computing resources, Software: Practice and Experience 50~(5) (2020) 558--595.

\bibitem{dai2019scheduling}
H.~Dai, X.~Zeng, Z.~Yu, T.~Wang, A scheduling algorithm for autonomous driving tasks on mobile edge computing servers, J. of Systems Arch. (2019).

\bibitem{ABURUKBA2020539}
R.~O. Aburukba, M.~Alikarrar, T.~Landolsi, K.~El-Fakih, Scheduling internet of things requests to minimize latency in hybrid fog–cloud computing, Future Generation Computer Systems 111 (2020) 539--551.

\bibitem{8941266}
J.~Meng, H.~Tan, X.-Y. Li, Z.~Han, B.~Li, Online deadline-aware task dispatching and scheduling in edge computing, IEEE Transactions on Parallel and Distributed Systems 31~(6) (2020) 1270--1286.

\bibitem{rao2021eco}
K.~Rao, G.~Coviello, W.-P. Hsiung, S.~Chakradhar, Eco: Edge-cloud optimization of 5g applications, in: IEEE/ACM International Symposium on Cluster, Cloud and Internet Computing (CCGrid), 2021.

\bibitem{zhang2022deadline}
Y.~Zhang, B.~Tang, J.~Luo, J.~Zhang, Deadline-aware dynamic task scheduling in edge--cloud collaborative computing, Electronics (2022).

\bibitem{8675170}
W.~Chen, B.~Liu, H.~Huang, S.~Guo, Z.~Zheng, When uav swarm meets edge-cloud computing: The qos perspective, IEEE Network 33 (2019).

\bibitem{postoaca2020deadline}
A.-V. Postoaca, C.~Negru, F.~Pop, Deadline-aware scheduling in cloud-fog-edge systems, in: IEEE/ACM International Symposium on Cluster, Cloud and Internet Computing (CCGRID), 2020.

\bibitem{fu2022kalmia}
Z.~Fu, J.~Ren, D.~Zhang, Y.~Zhou, Y.~Zhang, Kalmia: A heterogeneous qos-aware scheduling framework for dnn tasks on edge servers, in: IEEE Conference on Computer Communications (INFOCOM), 2022.

\bibitem{chen2021energy}
X.~Chen, J.~Zhang, B.~Lin, Z.~Chen, K.~Wolter, G.~Min, Energy-efficient offloading for dnn-based smart iot systems in cloud-edge environments, IEEE Transactions on Parallel and Distributed Systems (2022).

\bibitem{9996361}
S.~Shen, Y.~Ren, Y.~Ju, X.~Wang, W.~Wang, V.~C.~M. Leung, Edgematrix: A resource-redefined scheduling framework for sla-guaranteed multi-tier edge-cloud computing systems, IEEE Jour. on Sel. Areas in Comm. (2023).

\bibitem{LAI2020684}
P.~Lai, Q.~He, G.~Cui, X.~Xia, M.~Abdelrazek, F.~Chen, J.~Hosking, J.~Grundy, Y.~Yang, Qoe-aware user allocation in edge computing systems with dynamic qos, Future Generation Computer Systems 112 (2020).

\bibitem{LI201993}
C.~Li, J.~Bai, J.~Tang, Joint optimization of data placement and scheduling for improving user experience in edge computing, Journal of Parallel and Distributed Computing 125 (2019) 93--105.

\bibitem{9005248}
D.~Minovski, C.~Åhlund, K.~Mitra, Modeling quality of iot experience in autonomous vehicles, IEEE Internet of Things Journal 7~(5) (2020).

\bibitem{9714786}
L.~A.~b. Burhanuddin, X.~Liu, Y.~Deng, U.~Challita, A.~Zahemszky, Qoe optimization for live video streaming in uav-to-uav communications via deep reinforcement learning, IEEE Transactions on Vehicular Technology 71~(5) (2022) 5358--5370.

\bibitem{9191377}
P.~Vahidinia, B.~Farahani, F.~S. Aliee, Cold start in serverless computing: Current trends and mitigation strategies, in: International Conference on Omni-layer Intelligent Systems (COINS), 2020.

\bibitem{8917749}
Z.~Han, H.~Tan, X.-Y. Li, S.~H.-C. Jiang, Y.~Li, F.~C.~M. Lau, Ondisc: Online latency-sensitive job dispatching and scheduling in heterogeneous edge-clouds, IEEE/ACM Transactions on Networking (2019).

\bibitem{yolov8_ultralytics}
G.~Jocher, A.~Chaurasia, J.~Qiu, \href{https://github.com/ultralytics/ultralytics}{Ultralytics YOLOv8} (2023).

\bibitem{nagrath2006control}
I.~Nagrath, Control systems engineering, New Age International, 2006.

\bibitem{resnet18}
{NVIDIA AI IOT}, \href{https://github.com/NVIDIA-AI-IOT/trt_pose_hand}{Hand Pose Estimation And Classification} (2021).

\bibitem{vapnik1999nature}
V.~Vapnik, The nature of statistical learning theory, Springer science \& business media, 1999.

\bibitem{monodepth2github}
I.~Niantic, \href{https://github.com/nianticlabs/monodepth2}{Monodepth2: Monocular depth estimation from a single image} (2019).

\bibitem{ssd}
{AIZOOTech}, \href{https://github.com/AIZOOTech/FaceMaskDetection}{Detect faces and determine whether people are wearing mask} (2020).

\bibitem{raj2025ocularonebenchbenchmarkingdnnmodels}
S.~Raj, B.~A. Madhabhavi, K.~Astu, A.~A. Rajesh, P.~M, Y.~Simmhan, \href{https://arxiv.org/abs/2504.03709}{Ocularone-bench: Benchmarking dnn models on gpus to assist the visually impaired} (2025).
\newblock \href {http://arxiv.org/abs/2504.03709} {\path{arXiv:2504.03709}}.
\newline\urlprefix\url{https://arxiv.org/abs/2504.03709}

\bibitem{raj2025distanceestimationsupportassistive}
S.~Raj, B.~A. Madhabhavi, M.~Kumar, P.~Gupta, Y.~Simmhan, \href{https://arxiv.org/abs/2504.01988}{Distance estimation to support assistive drones for the visually impaired using robust calibration} (2025).
\newblock \href {http://arxiv.org/abs/2504.01988} {\path{arXiv:2504.01988}}.
\newline\urlprefix\url{https://arxiv.org/abs/2504.01988}

\bibitem{badiger2018violet}
S.~Badiger, S.~Baheti, Y.~Simmhan, Violet: A large-scale virtual environment for internet of things, in: Euro-Par: Parallel Processing, 2018.

\bibitem{10.1145/3492321.3519576}
I.~Gog, S.~Kalra, P.~Schafhalter, J.~E. Gonzalez, I.~Stoica, D3: a dynamic deadline-driven approach for building autonomous vehicles, in: Proceedings of 17th European Conference on Computer Systems, EuroSys, 2022.

\bibitem{shupeng2018safe}
L.~Shupeng, L.~Menglu, B.~M. Chen, Safe corridor based task interface for quadrotors, in: International Micro-Air Vehicles Conference, 2018.

\bibitem{bae2020self}
I.~Bae, J.~Moon, J.~Jhung, H.~Suk, T.~Kim, H.~Park, J.~Cha, J.~Kim, D.~Kim, S.~Kim, Self-driving like a human driver instead of a robocar: Personalized comfortable driving experience for autonomous vehicles, arXiv preprint arXiv:2001.03908 (2020).

\bibitem{gdpr_overview}
E.~Union, \href{https://gdpr-info.eu/}{General Data Protection Regulation (GDPR) Summary} (2018).

\end{thebibliography}
\balance

\clearpage
\nobalance
\appendix

\vspace{-0.3cm}
\section{\addc{Expected Execution Times for DNN Models}}\label{appendix:benchmarking}
\vspace{-0.2cm}
\addc{We provide additional details on the benchmark experiments we have run for each DNN model used in our evaluation (\S~\ref{sec:exp:setup}).}

\subsection{\addc{Edge Containers} } \label{appendix:e2e-time-edge}
\begin{figure}[t]
\centering
  \subfloat[{1 drone per edge}]{
    \includegraphics[width=0.4\columnwidth]{fgcs-figures/edge_Inf_plot.pdf}
   \label{fig:edge-benchmark-1drone}
  }
    \subfloat[{3 drones per edge}]{
\includegraphics[width=0.4\columnwidth]{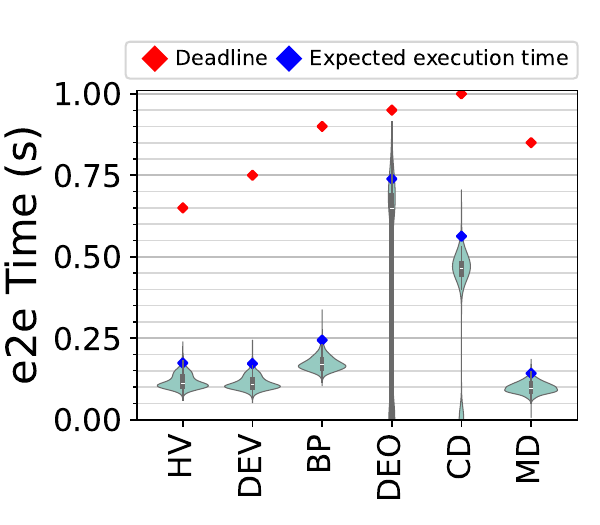}
    \label{fig:edge-benchmark-3drones}
  }
  \vspace{-0.1in}
\caption{\addc{DNN inferencing latency from benchmarks on Jetson Nano Edge Container}}
\label{fig:appendix:bm:edge}
\vspace{-0.2in}
\end{figure}

\addc{We run two scenarios for benchmarking of DNN models on our edge containers: with $1$ and $3$ client(s) (or drone) generating video feeds per container. For each experiment, we invoke $300$ gRPC synchronous calls for each DNN model per client per edge at 1 FPS using a container configured to emulate the performance of a Jetson Nano Edge. For both scenarios, the distribution of the total end-to-end execution duration of tasks executed for every DNN model across all edge containers on a host machine is reported in Figures~\ref{fig:edge-benchmark-1drone}, \ref{fig:edge-benchmark-3drones}.
The final expected execution duration on the edge ($t$) for the DNN model is then calculated as the average of the $99^{th}$ percentile of the total execution times for both scenarios and this value is reported in Table~\ref{setup-table} of \S~\ref{sec:exp:setup}.
}

\subsection{\addc{AWS Lambda Functions}\label{appendix:e2e-time-lambda}}
\begin{figure}[t]
\centering
  \subfloat[$7$ concurrent clients]{
   \includegraphics[width=0.33\columnwidth]{fgcs-figures/lambda_Inf_plot.pdf}
    \label{fig:lambdabenchmark-1drones-1VMs}
  }
    \subfloat[$21$ concurrent clients]{
   \includegraphics[width=0.33\columnwidth]{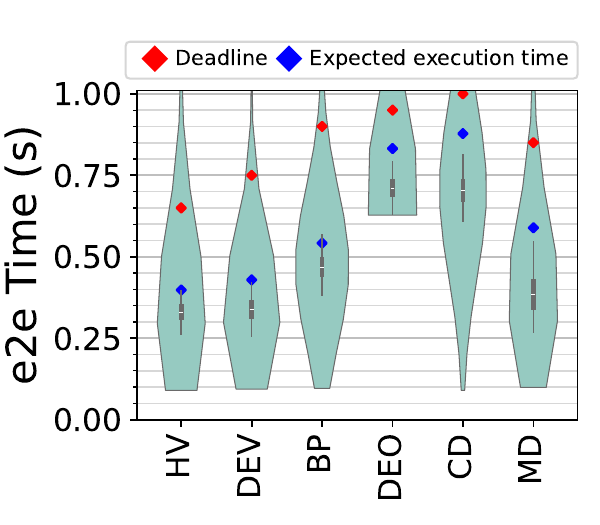}
    \label{fig:lambdabenchmark-1drones-3VMs}
  }
    \subfloat[$63$ concurrent clients]{
  \includegraphics[width=0.33\columnwidth]{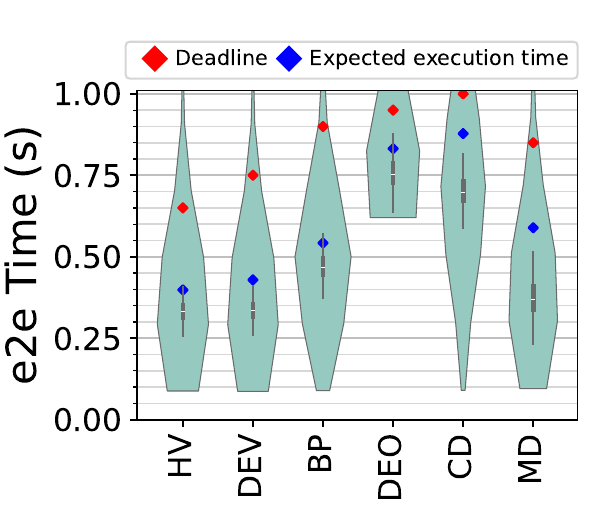}
    \label{fig:lambdabenchmark-3drones-3VMs}
  }
  \vspace{-0.1in}
\caption{\addc{DNN inferencing latency from benchmarks on AWS Lambda}}
\label{fig:appendix:bm-aws}
\vspace{-0.2in}
\end{figure}

\addc{We benchmark each DNN model using AWS Lambda cloud functions in $3$ scenarios that correspond to $7, 21$, and $63$ concurrent clients. For each scenario, we invoke $300$ HTTP requests for each DNN model per client at 1 FPS. The distribution of the total end-to-end execution duration of tasks executed for every DNN model across all clients is reported in Figures~\ref{fig:lambdabenchmark-1drones-1VMs}, \ref{fig:lambdabenchmark-1drones-3VMs} and \ref{fig:lambdabenchmark-3drones-3VMs}.
The final expected execution duration on cloud ($\hat{t}$) is then calculated as the average of the $95^{th}$ percentile of the total execution duration for all three scenarios and is set as in Table~\ref{setup-table} (\S~\ref{sec:exp:setup}).}

\addcr{Since edge inference times are tighter than AWS FaaS times, using the $99^{th}$ percentile on the edge accounts for nearly all task completions, minimizing outliers. In contrast, the cloud's higher variability and more outliers require a less stringent $95^{th}$ percentile threshold.}

\vspace{-0.3cm}
\section{\addc{Normalized Costing for DNN Models}}\label{appendix:costing}
\vspace{-0.2cm}
\addc{We provide additional details on the realistic costing for executing each DNN model used in our evaluation (\S~\ref{sec:exp:setup}).}

\subsection{\addc{Edge Containers}}

\addc{We calculate the absolute operational cost on the edge per time unit by summing up the cost of the edge device, $INR~10,000$ ($\approx US\$120$) for a Jetson Nano and $INR~55,000$ ($\approx US\$600$) for Jetson Orin Nano, and the cost of electricity consumed over an estimated lifespan of 3 years for the edge device. Using the actual execution (inferencing) duration from the edge benchmark experiments~\ref{appendix:e2e-time-edge}, we calculate the amortized operational cost of execution on the edge for all DNN models for each execution. Further, we normalize the cost values obtained by scaling them with respect to the minimum cost value obtained across all DNN models and set this as \modcr{$\mathcal{K}$} in Table~\ref{setup-table} (\S~\ref{sec:exp:setup}).}

\subsection{\addc{AWS Lambda Functions}}
\addc{AWS Lambda pre-specifies unit cost of execution in GB-seconds~\footnote{\url{https://aws.amazon.com/lambda/pricing/}}, which varies with the memory allocated to the Lambda function and the region being used. These are provided as lambda function configurations in our experiments (\S~\ref{sec:exp:setup}). Again, we make use of the benchmark experiments described in~\ref{appendix:e2e-time-lambda} and consider only the inference latency. We calculate the expected inferencing time latency as the average of the median of inferencing latency values for all three scenarios and multiply it by the unit cost of execution for each function in the Asia Pacific (Mumbai) AWS region, which is INR~$0.000001397$ per GB-second.
This gives us a total cost of execution on the lambda for all DNN models. Finally, we obtain the normalized cost values by scaling them with respect to the minimum cost value obtained across all DNN models and set them as \modcr{$\hat{\mathcal{K}}$} in Table~\ref{setup-table} (\S~\ref{sec:exp:setup}).
}

\vspace{-0.3cm}
\section{\addcr{Adapting to Network Variability under Varying Workloads}\label{sec:appendix-network-variability}}
\vspace{-0.2cm}
\begin{figure}[!t]
  \centering
\subfloat[Latency Variability]{
    \includegraphics[width=0.4\columnwidth]{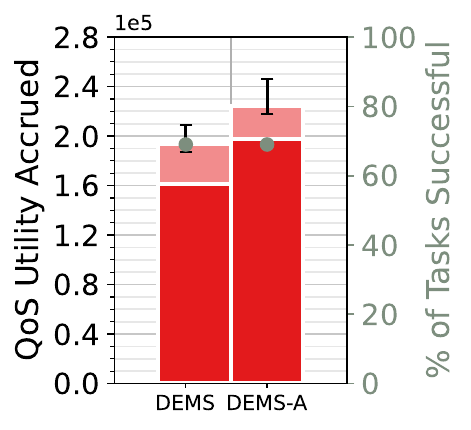}
   \label{fig:nw-adapt:lat-util:lat:appendix}
  }\qquad
  \subfloat[Bandwidth Variability]{
   \includegraphics[width=0.4\columnwidth]{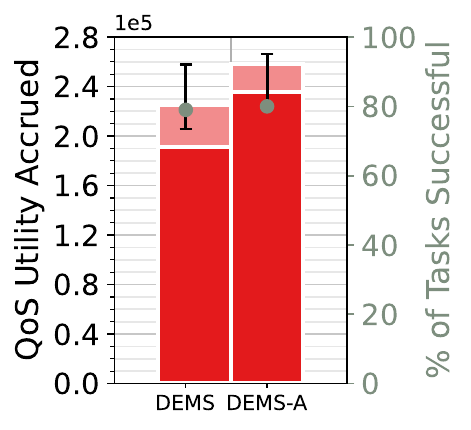}
    \label{fig:nw-adapt:lat-util:bw:appendix}
  }
  \vspace{-0.07in}
	\caption{\addcr{Benefits of \DS-A over \DS for \threeBP in the presence of \textit{latency (left)} and \textit{bandwidth (right) variability}.The lighter-shaded stack indicates tasks completed on the cloud, while the darker shade are tasks completed on the edge.}}
    \label{fig:nw-adapt:lat-util:3dp}
    \vspace{-0.1in}
\end{figure}

\begin{figure}[t]
\centering

\begin{tabular}{@{}c@{}|@{}c@{}}
  \subfloat[\DS: Latency Variability]{
     \includegraphics[width=0.49\columnwidth]{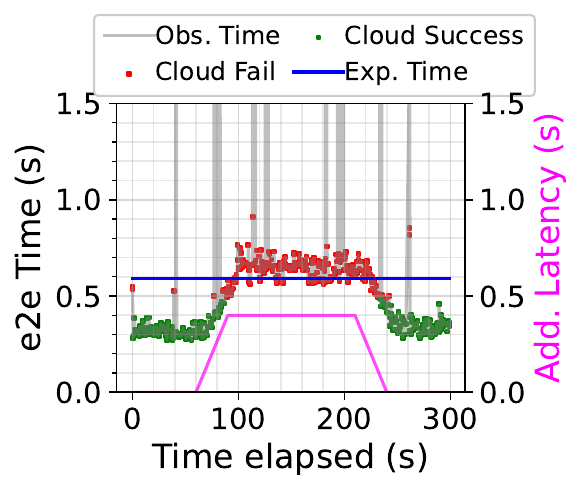}
   \label{fig:nw-adapt:lat:dems:appendix}
  }&
  \subfloat[\DS: Bandwidth Variability]{
     \includegraphics[width=0.48\columnwidth]{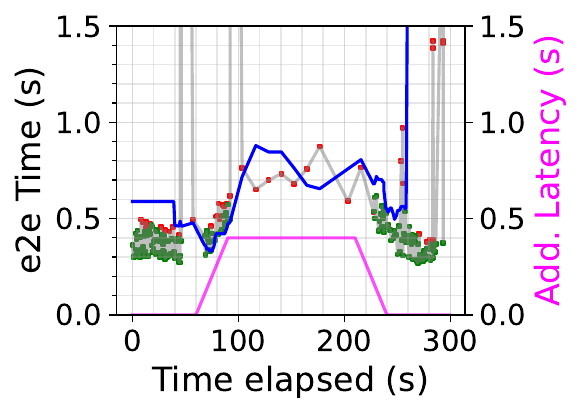}
   \label{fig:nw-adapt:bw:dems:appendix}
  }\\
  \subfloat[\DS-A: Latency Variability]{
   \includegraphics[width=0.49\columnwidth]{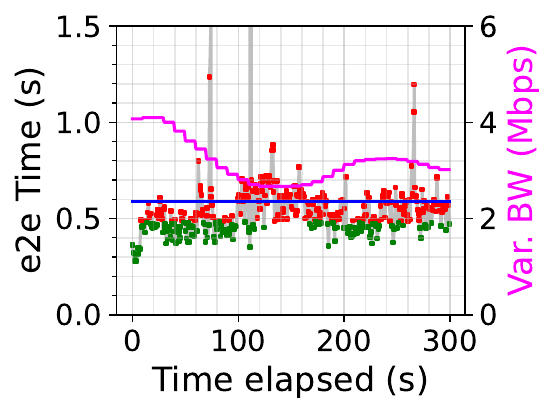}
    \label{fig:nw-adapt:lat:demsa:appendix}
  }&
  \subfloat[\DS-A: Bandwidth Variability]{
   \includegraphics[width=0.48\columnwidth]{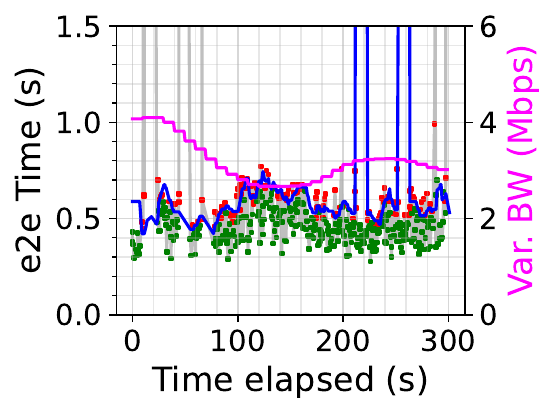}
    \label{fig:nw-adapt:bw:demsa:appendix}
  }
\end{tabular}
\vspace{-0.07in}
\caption{\addcr{End-to-end latency for \threeBP with \textit{\DS (top)} and \textit{\DS-A (bottom)} heuristics in the presence of \textit{variability in latency (left)} and \textit{bandwidth (right)}.}}
\label{fig:nw-adapt:3dp}
  \vspace{-0.1in}
\end{figure}

\addcr{We report the effectiveness of \DS-A algorithm over \DS on \threeBP workload in the presence of latency and bandwidth variability in Fig.~\ref{fig:nw-adapt:lat-util:3dp},\ref{fig:nw-adapt:3dp}. The setup of the experiment is same as described in Sec.~\ref{sec:exp:nw} and the plots demonstrate same parameters. In the case of \threeBP workload, we observe that \DS-A is able to improve the utility by $16\%$ for latency variability and by $15\%$ for bandwidth variability, while completing a similar number of on-time tasks.}

\vspace{-0.3cm}
\section{\addcr{Plots with \# of Tasks Successful}} \label{appendix:sec:absolute-plots}
\vspace{-0.2cm}
\begin{figure}[!t]
 \centering
 \includegraphics[width=0.9\columnwidth]{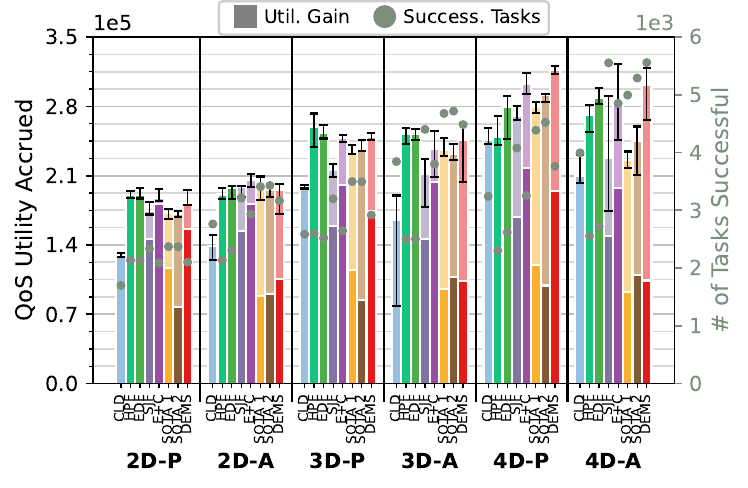}
  \vspace{-0.1in}
    \caption{\addcr{Reference plot from Fig.~\ref{fig:baseline:bars} representing \# of tasks successful.}}
    \label{fig:baseline:bars:appendix}
    \vspace{-0.1in}
\end{figure}

\begin{figure}[!t]
 \centering
 \includegraphics[width=0.7\columnwidth]{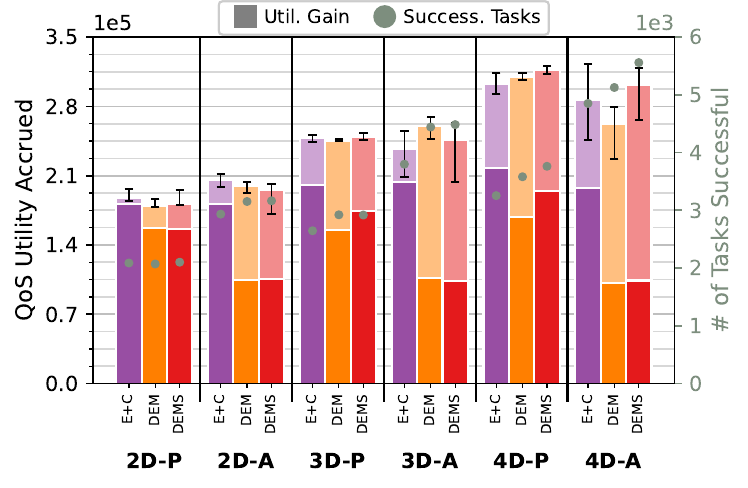}
  \vspace{-0.1in}
    \caption{\addcr{Reference plot from Fig.~\ref{fig:migration-with-stealing} representing \# of tasks successful.}}
    \label{fig:baseline:dem-dems:appendix}
    \vspace{-0.2in}
\end{figure}

\begin{figure}[!t]
  \centering
\subfloat[Latency Variability]{
    \includegraphics[width=0.3\columnwidth]{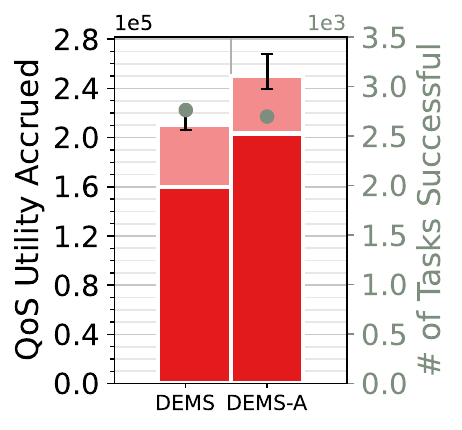}
   \label{fig:nw-adapt:lat-util:lat:appendix:4dp}
  }\qquad
  \subfloat[Bandwidth Variability]{
   \includegraphics[width=0.3\columnwidth]{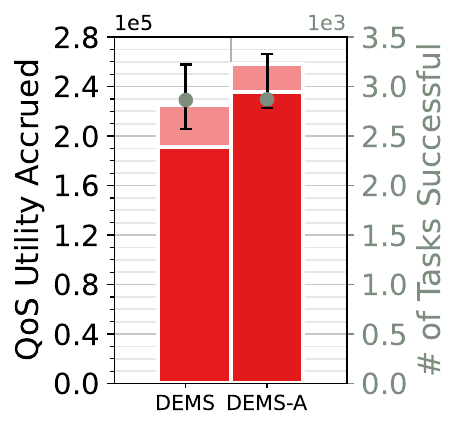}
    \label{fig:nw-adapt:lat-util:bw:appendix:4dp}
  }
  \vspace{-0.1in}
	\caption{\addcr{Reference plot from Fig.~\ref{fig:nw-adapt:lat-util} representing \# of tasks successful.}}
    \label{fig:nw-adapt:lat-util:appendix}
    \vspace{-0.2in}
\end{figure}

\begin{figure}[!t]
  \centering
\subfloat[Latency Variability]{
    \includegraphics[width=0.3\columnwidth]{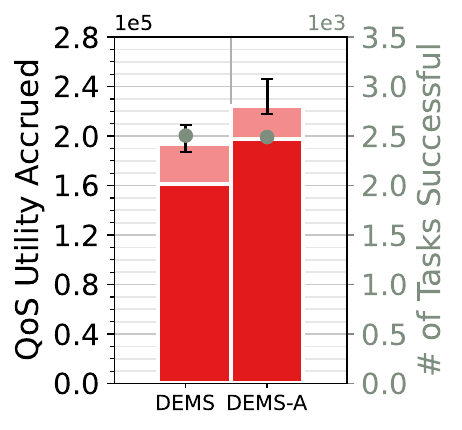}
   \label{fig:nw-adapt:lat-util:lat:appendix:3dp}
  }\qquad
  \subfloat[Bandwidth Variability]{
   \includegraphics[width=0.3\columnwidth]{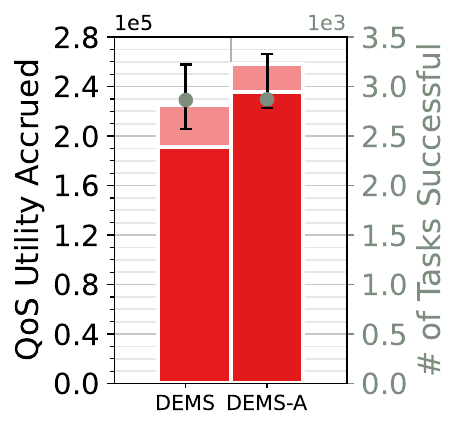}
    \label{fig:nw-adapt:lat-util:bw:appendix:3dp}
  }
  \vspace{-0.1in}
	\caption{\addcr{Reference plot from Fig.~\ref{fig:nw-adapt:lat-util:3dp} representing \# of tasks successful.}}
    \label{fig:nw-adapt:lat-util:appendix:3dp}
    \vspace{-0.1in}
\end{figure}

\begin{figure}[!t]
  \centering
	\includegraphics[width=0.5\columnwidth]{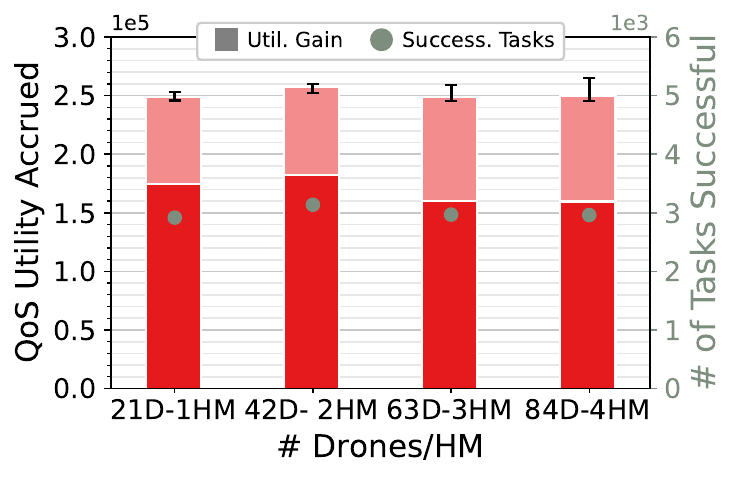}
 \vspace{-0.1in}
	\caption{\modcr{Reference plot from Fig.~\ref{fig:weakScaling} representing \# of tasks successful.}}
    \label{fig:weakScaling:appendix}
    \vspace{-0.2in}
\end{figure}

\addcr{We include Fig.~\ref{fig:baseline:bars:appendix}, \ref{fig:baseline:dem-dems:appendix}, \ref{fig:nw-adapt:lat-util:appendix}, \ref{fig:nw-adapt:lat-util:appendix:3dp}, and \ref{fig:weakScaling:appendix} to present the absolute values of successful tasks. This helps readers better relate these figures to the percentage of tasks completed, providing a clearer understanding of task success rates.}

\vspace{-0.3cm}
\section{\addc{Hardware Used for Practical Validation}}
\vspace{-0.2cm}
\begin{figure}[!t]
\centering%
\subfloat[Tello Drone]{
    \includegraphics[width=0.4\columnwidth]{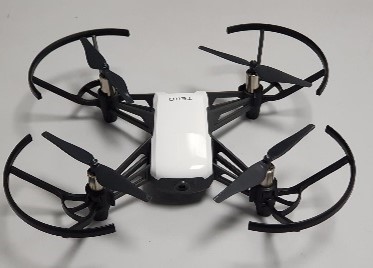}
   \label{fig:tello}
  }
  \subfloat[Jetson Orin Nano]{
   \includegraphics[width=0.35\columnwidth]{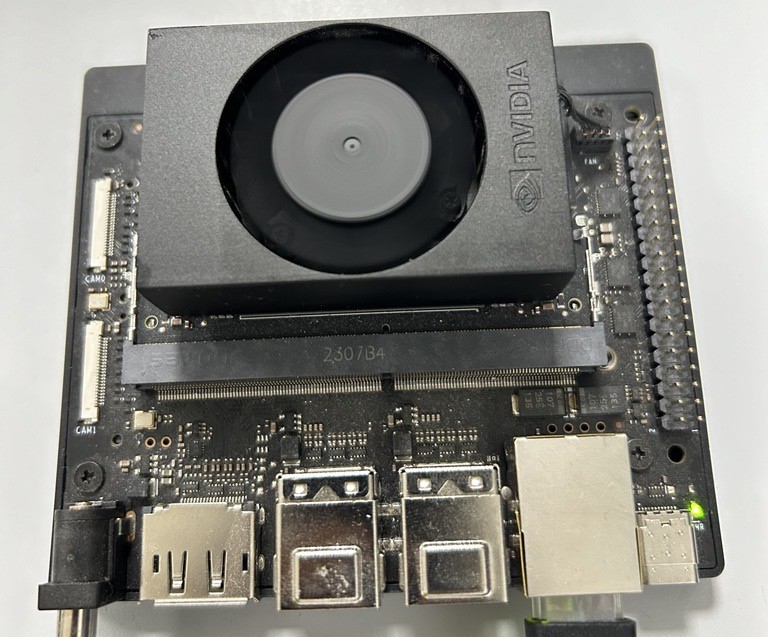}
    \label{fig:orin-nano}
  }
  \vspace{-0.1in}
\caption{\addc{Hardware Components Required for the Experimental Setup}}
    \label{fig:hardware-components}
    \vspace{-0.15in}
\end{figure}

\addc{As described in \S~\ref{subsec:hardware-validation}, we perform a field validation for the VIP application using our proposed scheduling algorithms with real hardware. Figure~\ref{fig:tello} shows the Tello Drone used in the experiments while Figure~\ref{fig:orin-nano} shows the Nvidia Jetson Orin Nano Developer Kit used as the edge accelerator.}


\section{\addc{Supplementary Video}}\label{sec:supplementary-video}
\addc{The videos demonstrating our field validation of the scheduling heuristics for the VIP application domain within our university campus outdoor setting(\S~\ref{subsec:hardware-validation}) is \href{https://drive.google.com/file/d/1qAhXa_NEZlMclfevG0aRnnQGDmbDvGoW/view?usp=sharing}{available at this link}.}

\end{document}